\documentclass[10pt,epsfig,A4]{article}
\usepackage{epsfig}
\usepackage{amsmath}
\usepackage{amsthm}
\usepackage{color}
\usepackage{subfigure}
\usepackage{amssymb}
\usepackage{fancyhdr}
\usepackage{enumerate}
\usepackage{mathrsfs}
\usepackage{dsfont}
\usepackage{stmaryrd}

\renewcommand{\baselinestretch}{1}
\newtheorem{lemma}{Lemma}

\newtheorem{theorem}{Theorem}

\theoremstyle{definition}

\newtheorem*{example*}{Example}
\newtheorem*{remark*}{Remark}
\newcommand{\bre}{\begin{equation}}
\newcommand{\ere}{\end{equation}}
\newcommand{\be}\[
\newcommand{\ee}\]
\newcommand{\bra}{\begin{eqnarray}}
\newcommand{\era}{\end{eqnarray}}
\newcommand{\ba}{\begin{eqnarray*}}
\newcommand{\ea}{\end{eqnarray*}}
\newcommand{\bfg}{\begin{figure}[hbtp]}
\newcommand{\efg}{\end{figure}}
\newcommand{\bver}{\begin{verbatim}}
\newcommand{\ever}{\end{verbatim}}
\newcommand{\bit}{\begin{itemize}}
\newcommand{\eit}{\end{itemize}}
\newcommand{\ben}{\begin{enumerate}}
\newcommand{\een}{\end{enumerate}}

\def\argsup{\mathop{\rm argsup}}

\newcommand{\comment}[1]{}

\renewcommand{\thesection}{\Roman{section}}

\newlength{\tmpbigbar}

\newcommand{\vect}[1]
{\mbox{\boldmath $#1$}}

\newcommand{\decision}[2]{\;\raisebox{-0.4ex}{$\stackrel{#1}{\stackrel{\lessgtr}{\raisebox{-1.5ex}{$\scriptstyle #2$}}}$}\;}

\newfont{\boldlarge}{msbm10 scaled 1100}
\newfont{\boldsmall}{msbm10 scaled 800}
\newcommand{\RealF}{\mbox{\boldlarge R}}

\newcommand{\NaturalF}{\mbox{\boldlarge N}}

\newcommand{\Expt}{\mbox{\boldlarge E}}
\newcommand{\Prob}{\mbox{\boldlarge P}}
\newcommand{\defn}{ \stackrel{\triangle}{=} }
\newcommand{\dfn}{\triangleq}

\newcommand{\Simplex}{\mbox{\boldlarge S}}
\newcommand{\Linf}[1]{\left\|#1\right\|_\infty}
\newcommand{\brk}[1]{\left\langle #1\right\rangle}

\newcommand{\m}[1]
{\mathcal{#1}}
\newcommand{\sst}{\scriptscriptstyle}
\newcommand{\sstm}[1]
{\sst\mathcal{#1}}

\def\eps{\varepsilon}

\def\bigo{{\mathrm{O}}}
\def\lito{\mathrm{o}}

\newcommand{\samp}{\hspace{-0.12cm}\downarrow\hspace{-0.05cm}}

\def\ind{\mathds{1}}
\def\wt{\widetilde}

\begin{document}

\title{
{\bf {Achieving the Empirical Capacity Using Feedback Part I:
Memoryless Additive Models}}}

\author{ {\bf Ofer Shayevitz and Meir Feder}\\
         {\sl Dept. of Electrical Engineering Systems}\\
         {\sl Tel Aviv University}\\
         {\sl Tel Aviv 69978, Israel}\\
         {\sl Email: ofersha@eng.tau.ac.il, meir@eng.tau.ac.il} }

\date{ }

\maketitle

\vspace{0.5cm}


\pagestyle{fancy}



\thispagestyle{empty}

\begin{abstract}
We address the problem of universal communications over an unknown
channel with an instantaneous noiseless feedback, and show how
rates corresponding to the empirical behavior of the channel can
be attained, although no rate can be guaranteed in advance. First,
we consider a discrete modulo-additive channel with alphabet
$\m{X}$, where the noise sequence $Z^n$ is \textit{arbitrary and
unknown} and may causally depend on the transmitted and received
sequences and on the encoder's message, possibly in an adversarial
fashion. Although the classical capacity of this channel is zero,
we show that rates approaching the \textit{empirical capacity}
$\log{|\m{X}|}-H_{\rm emp}(Z^n)$ can be universally attained,
where $H_{\rm emp}(Z^n)$ is the empirical entropy of $Z^n$. For
the more general setting where the channel can map its input to an
output in an arbitrary unknown fashion subject only to causality,
we model the empirical channel actions as the modulo-addition of a
realized noise sequence, and show that the same result applies if
common randomness is available. The results are proved
constructively, by providing a simple sequential transmission
scheme approaching the empirical capacity. In part II of this work
we demonstrate how even higher rates can be attained by using more
elaborate models for channel actions, and by utilizing possible
empirical dependencies in its behavior.
\end{abstract}


{\renewcommand{\thefootnote}{} \footnotetext{O. Shayevitz is
supported by the Adams Fellowship Program of the Israel Academy of
Sciences and Humanities. This research was supported in part by
the Israel Science Foundation, grant no. 223/05.}}

\newpage

\section{Introduction}\label{sec:introduction}
The capacity of a channel is classically defined as the supremum
of all rates for which communication with arbitrarily low
probability of error can be guaranteed in advance. However, when a
noiseless feedback link between the receiver and the transmitter
exist, one does not necessarily have to commit to a rate prior to
transmission, and communication can take place using some
sequential scheme at a variable rate determined by the specific
realization of the channel, thus the better the channel
realization the higher the rate of transmission. When the channel
law is known this approach cannot yield average rates exceeding
those attainable by fixed rate feedback schemes, and for large
classes of channels cannot even exceed the rates of non-feedback
schemes
\cite{shannon_zero_error}\cite{feedback_additive_noise}\cite{capacity_feedback_memoryless}.
The variable-rate approach may, however, have the advantages of a
better error exponent and a lower complexity. Several transmission
schemes possessing such merits were proposed for the binary
symmetric channel (BSC) \cite{horstein}\cite{Schalkwijk-Post}, the
Gaussian additive noise channel
\cite{Schalkwijk-Kailath}\cite{yamamoto}, discrete memoryless
channels (DMC) \cite{Horstein-report}\cite{yamamoto} and
finite-state channels (FSC) \cite{Ooi-Wornell}.

When the channel law is unknown to some degree, variable rate
feedback schemes become even more attractive, as the realized
channel may sometimes be explicitly or implicitly estimated via
feedback. In \cite{nadav_phd}\cite{nadav_sorrento}, a rate
universal scheme for unknown DMC with a random decision time was
suggested (later termed \textit{rateless coding}), attaining a
rate equal to the mutual information of the channel in use for any
selected input distribution. Following this lead, a universal
variable rate transmission schemes for compound BSC and Z-channels
with feedback was introduced \cite{Tchamkerten-Telatar}, and shown
to attain any fraction of the realized channel's capacity and
achieve the Burnashev error exponent \cite{burnashev_exp}. In
\cite{feedback-thesis_ooi} it was shown that for compound FSC with
feedback, it is possible to transmit at a rate approaching the
mutual information of the realized channel for any Markov input
distribution, by an incremental universal compression of the
errors, e.g. via Lempel-Ziv coding.

So far, however, the variable-rate approach was not applied to
more stringent channel uncertainty models, where the channel
behavior is arbitrary or even adversarial. As a motivating
example, consider a binary modulo-additive channel with feedback,
where the noise sequence is an individual sequence (i.e.,
deterministic and unknown). Let us assume (at first) that the
fraction of '1's in the noise sequence (namely the fraction of
errors inserted by the channel) is a-priori known to be
$p\in[0,1]$ at the most. The fixed rate communication problem in
this setting has been addressed before in several different
contexts. In a classical work \cite{Berlekamp-Thesis}, Berlekamp
considered this model in the context of error correction
capability with feedback, where the receiver is required to
correct the errors inserted by the channel and uniquely recover
the transmitted message. Since no decoding errors are allowed, the
noise sequence in this case can also be thought of as being
generated by an adversary that knows the message and the coding
scheme, but is ``power limited'' by $p$. Berlekamp showed that
whenever $p\geq \frac{1}{3}$, there exists an adversarial strategy
for error insertion such that the receiver cannot hope to separate
even three messages, and so the capacity is zero. For smaller $p$,
he was able to show that the capacity is upper bounded
by\footnote{$h_{\sst B}(\cdot)$ is the binary entropy function.}
$1-h_{\sst B}(p)$ and a (tight) straight line tangent to
$1-h_{\sst B}(p)$, intersecting the horizontal axis at
$p=\frac{1}{3}$. The convex part of this bound was later shown to
be tight as well \cite{Zigangirov76}.

The same communication problem can also be studied in the context
of the (discrete memoryless) Arbitrarily Varying Channel (AVC). In
an AVC setting, a memoryless channel law is selected from a given
set (state space) at each time point, in an arbitrary unknown
manner. The AVC without feedback was studied extensively
\cite{blackwell-breiman-thomasian-1960}\cite{ahlswede_dichotomy}\cite{capacity_AVC_constraints},
and shown to yield different capacities depending on the error
criterion (average/maximum error probability) and also on the
existence of common randomness (resulting in the so-called
\textit{random-code} capacity, which is the same under both error
criteria). Within the AVC framework, the channel under discussion
is a binary AVC with two states, a clean channel and an inverting
channel, where the noise sequence becomes the state sequence and
the maximal fraction of channel errors $p$ yields a state
constraint. Under the maximum error probability criterion, this
AVC with feedback is equivalent to Berlekamp's setting, the
capacity of which was given above. Interestingly, it turns out
that even \textit{without feedback}, the \textit{random-code}
capacity of this channel is given by $1-h_{\sst B}(p)$ for any
$p<\frac{1}{2}$ (and zero otherwise)\footnote{In fact, this is
also the deterministic coding capacity without feedback, under the
\textit{average} error probability criterion.} and can be attained
with merely $\Omega(\log{n})$ bits of common randomness
\cite{ahlswede_dichotomy}\cite{private_codes}. This small amount
of randomness can be generated via feedback with a negligible
impact on rate, hence the capacity of the discussed binary channel
with feedback coincides with its (non-feedback) random-code
capacity, yielding a significant gain relative to Berlekamp's
capacity through the use of randomness. This approach of extending
non-feedback AVC results to the feedback regime was also taken in
\cite{cr_it} (albeit without state constraints) where it was shown
that the feedback capacity of an AVC is equal to its random-code
capacity.

Berlekamp's result and the AVC approach are limited by the
requirement to commit to a fixed rate prior to transmission, and
so positive rates are obtained only under noise/state sequence
constraints. As we shall see, variable-rate coding can be used to
obtain a much stronger result that applies to any noise sequence
without any constraint (i.e., $p=1$). As a corollary of our main
result, we constructively show that for the binary channel under
discussion, rates arbitrarily close to $1-h_{\sst B}(p_{\rm emp})$
can be achieved by a simple (deterministic, algorithmic)
sequential feedback scheme with probability approaching one and a
vanishing (maximum) error probability\footnote{The probabilities
in this case are taken w.r.t. randomness created via the feedback
link.}, where $p_{\rm emp}$ is the \textit{empirical} fraction of
'1's in the individual noise sequence. Thus, although the
fixed-rate capacity is zero when there are no constraints on the
noise sequence, one can opportunistically attain rates approaching
what would have been the capacity of the channel had the fraction
of '1's in the noise sequence been known in advance. It is
therefore only appropriate to call the quantity $1-h_{\sst
B}(p_{\rm emp})$ the (zero-order, modulo-additive)
\textit{empirical capacity} of the realized channel.

More generally, in this paper we consider a discrete channel with
feedback over a common input/output alphabet $\m{X}$, that maps
its input to an output in a modulo-additive fashion, where the
corresponding noise sequence is \textit{arbitrary and unknown} and
may causality depend on the transmitted and received sequences and
on the encoder's message. We constructively show that rates
arbitrarily close to the (zero-order, modulo-additive) empirical
capacity $\log{|\m{X}|}-H_{\rm emp}(Z^n)$ can be achieved by a
simple sequential scheme with probability approaching one, where
$H_{\rm emp}(Z^n)$ is the (zero-order) empirical entropy of the
noise sequence $Z^n$. Furthermore, we consider the more general
setting where the channel can map its input to an output in an
\textit{arbitrary unknown fashion} (not necessarily
modulo-additive), subject only to causality. By modelling the
channel actions as the modulo-addition of a realized noise
sequence, we show that the corresponding empirical capacity can be
achieved, if common randomness is allowed. These channel models
can also be interpreted as adversarial, where an adversary
(jammer) that knows the transmission scheme and is in possession
of the transmitted message, causally listens to the transmitted
and received sequences and employs an arbitrary unknown jamming
strategy.

The paper is organized as follows. In section \ref{sec:prelim}
some notations and useful Lemmas are given. The channel model and
the main result of the paper are provided in section
\ref{sec:result}. A finite-horizon feedback transmission scheme
achieving the empirical capacity in a modulo-additive setting is
described in section \ref{sec:scheme}, and its analysis appears in
section \ref{sec:analysis}. A short discussion is provided in
section \ref{sec:summary}. The horizon-free variant of the scheme
appears in Appendix \ref{app:horizon_free}, and its extension to
general causal channels under the modulo-additive model using
common randomness, is discussed in Appendix \ref{app:dithering}.
Part II of this work \cite{empirical_capacity_partII} is dedicated
to the investigation of more elaborate models for channel actions,
and the corresponding gain in rate that may be achieved by
utilizing empirical dependencies in the channel's behavior.

\section{Notations and Preliminaries}\label{sec:prelim}
The following standard asymptotic notations are used:
\begin{align*}
f(n)=\bigo(g(n)) &\;\Longleftrightarrow\;
\limsup_{n\rightarrow\infty} \left|\frac{f(n)}{g(n)}\right| <
\infty
\\
f(n)=\lito(g(n)) &\;\Longleftrightarrow\;
\lim_{n\rightarrow\infty} \frac{f(n)}{g(n)} = 0
\\
f(n)=\Omega(g(n)) &\;\Longleftrightarrow\;
\liminf_{n\rightarrow\infty} \left|\frac{f(n)}{g(n)}\right| > 0
\\
f(n)=\omega(g(n)) &\;\Longleftrightarrow\;
\lim_{n\rightarrow\infty} \frac{f(n)}{g(n)} =\infty
\end{align*}
For $n\in\NaturalF$ we use the convention
$\brk{\hspace{0.02cm}n}\dfn \left\{0,1,\ldots,n-1\right\}$. All
logarithms are taken to the base of 2. For vectors, we write
$z_m^n=(z_m,z_{m+1}\ldots,z_n)$ which by convention is the null
string if $m>n$, and use $z^n=z_1^n$ for short. For real valued
vectors, $\|\cdot\|_\infty$ is the $\m{L}_\infty$ norm. Random
variable (r.v's) are usually denoted by uppercase letters, with
the corresponding lower-case letters for realizations. We write
$H(\cdot)$ for the entropy function, $h_{\sst B}(\cdot)$ for the
binary entropy function, and $D(\cdot\|\cdot)$ for relative
entropy. A finite alphabet $\m{X}$ in this paper is taken to be
the set $\m{X}=\brk{|\m{X}|}$ associated with the modulo-addition
operator $+$, unless otherwise stated.
\begin{lemma}[Entropy $\m{L}_\infty$ bound]\label{lem:entopry_Linf_bound}
Let $\vect{p}$ be a probability distribution over a finite
alphabet $\m{X}$. Then
\begin{equation*}
H(\vect{p}) \;\geq\; \log{|\m{X}|}\Big{(}1 -
|\m{X}|\hspace{0.02cm}\big{\|}\vect{p}-\vect{p}_u\big{\|}_\infty\Big{)}
\end{equation*}
where $\vect{p}_u$ is the uniform distribution over $\m{X}$.
\end{lemma}
\begin{proof}
See Appendix \ref{app:lemmas}.
\end{proof}

For a sequence $z^n\in\m{X}^n$, the number of occurrences of the
symbol $i\in\m{X}$ is denoted by $n_i(z^n)$. The \textit{empirical
distribution} of $z^n$ is the vector of relative symbol
occurrences in $z^n$,
\begin{equation*}
\vect{p}_{\rm emp}(z^n) \;\dfn\;\left(\frac{n_{\sst
0}(z^n)}{n}\,,\frac{n_{\sst
1}(z^n)}{n}\,,\ldots,\,\frac{n_{|\sstm{X}|-1}(z^n)}{n}\right)
\end{equation*}
where by convention, the empirical distribution of a null string
is taken to be uniform. When $z^n$ is a binary sequence, we write
$p_{emp}(z^n)$ for its empirical fractions of '1's, and loosely
refer to $p_{emp}(z^n)$ as the empirical distribution of $z^n$.
The (zero-order) \textit{empirical entropy} of $z^n$ is
$H(\vect{p}_{\rm emp}(z^n))$, the entropy pertaining to the
empirical distribution, and is denoted by $H_{\rm emp}(z^n)$ for
short. For a binary sequence, the empirical entropy is written
$h_{\sst B}(p_{emp}(z^n))$.

A \textit{sequential probability estimator} over a finite alphabet
$\m{X}$ is a sequence of nonnegative functions
$\left\{\widehat{p}_k(\cdot|z^{k-1})\right\}_{k=1}^\infty$ which
sum to unity for any $k\in\NaturalF\,,z^{k-1}\in\m{X}^{k-1}$. As
usual, the function $\widehat{p}_k(\cdot|z^{k-1})$ is thought of
as a \textit{probability assignment} for the next symbol $z_k$
given past observations $z^{k-1}$. The probability assigned by the
sequential estimator to any finite individual sequence $z^{n}$ is
therefore defined as
\begin{equation*}
\widehat{p}\,(z^n) \;\dfn\;
\prod_{k=1}^n\widehat{p}_k(z_k|z^{k-1}) \;,\qquad z^n\in\m{X}^n
\end{equation*}
We would also be interested in the following quantity,
\begin{equation*}
\widehat{p}\,(z^n\|w^n) \;\dfn\;
\prod_{k=1}^n\widehat{p}_k(z_k|w^{k-1}) \;,\qquad
z^n,w^n\in\m{X}^n
\end{equation*}
which is the probability assigned to the individual sequence $z^n$
by a sequential estimator matched to a different individual
sequence $w^n$, namely the case of sequential estimation from
noisy observations.

A well known probability estimator is the Krichevsky-Trofimov (KT)
estimator \cite{Krichevsky-Trofimov} given by
\begin{equation*}
\widehat{p}_k^{\;\sst{KT}}(i|z^{k-1}) =
\frac{n_i(z^{k-1})+\frac{1}{2}}{k-1+\frac{|\m{X}|}{2}}\;,\qquad
i\in\m{X}
\end{equation*}
The following Lemma shows that the per-symbol codelength assigned
by the KT estimator to any individual sequence is close to its
empirical entropy.
\begin{lemma}[KT redundancy
\cite{universal_sequential_coding_single_message}]\label{lem:KT}
For any individual sequence $z^n\in\m{X}^n$,
\begin{equation*}
-\frac{1}{n}\log\widehat{p}^{\;\sst{KT}}(z^n) \leq H_{\rm
emp}(z^n) +\frac{(|\m{X}|-1)\log{n}+\bigo(1)}{2n}
\end{equation*}
\end{lemma}

A sequential probability estimator is said to be a \textit{KT($b$)
estimator} if it can be obtained from a KT estimator by updating
the estimates \textit{at least} once per $b$ symbols. Such an
estimator is given by
\begin{equation*}
\widehat{p}_k^{\;\sst{KT(b)}}(i|z^{k-1}) =
\widehat{p}_k^{\;\sst{KT}}(i|z^{\nu_k}) \;,\qquad i\in\m{X}
\end{equation*}
where $\{\nu_k\}_{k=1}^\infty$ is a nondecreasing index sequence
determining the positions where the KT estimates are updated, thus
satisfying $k-b\leq \nu_k \leq k-1$. In the sequel, we will be
interested in the excess redundancy incurred when using a KT($b$)
estimator in lieu of the KT estimator, and possibly when the
estimator is matched to a different individual sequence, i.e., the
case of noisy observations.
\begin{lemma}[Noisy KT($b$) excess redundancy]\label{lem:KTb}
Let $\widehat{p}^{\;\sst{KT(b)}}$ be a KT($b$) estimator for some
$b\in\NaturalF$. Then for any pair of individual sequences
$z^n,w^n\in\m{X}^n$,
\begin{equation}\label{eq:rate_nonbinary1}
-\log\frac{\widehat{p}^{\;\sst{KT(b)}}(z^n\|w^n)}{\widehat{p}^{\;\sst{KT}}(z^n)}
\leq 2|\m{X}|\big{(}b+d(z^n,w^n)-1\big{)}\log{2ne}
\end{equation}
where $d(\cdot,\cdot)$ is the Hamming distance operator.
\end{lemma}
\begin{proof}
See Appendix \ref{app:lemmas}.
\end{proof}

Let $z^n\in\m{X}^n$, $b^n\in\{0,1\}^n$, and let $\sigma_k(b^n)$ be
the index of the $k$th nonzero element in $b^n$. Define $z^n\samp
b^n\in\m{X}^{n_1(b^n)}$ to be the vector whose $k$th element is
$z_{\sigma_k(b^n)}$, i.e., $z^n\samp b^n$ is a sample of size
$n_1(b^n)$ of $z^n$ where sampling positions are determined by the
'1's in $b^n$. The next Lemma bounds the probability that the
deviation (measured in the $\m{L}_\infty$ norm) between the
empirical distributions of a r.v. sequence, and that of a
fixed-size random sample without replacement from that sequence,
exceeds some threshold. It is a direct consequence of a result by
Hoeffding \cite{hoeffding}
\begin{lemma}[Sampling without replacement]\label{lem:hoeffding_sampling}
Let $Z^n,B^n$ be a pair of statistically independent r.v.
sequences, where $Z^n$ takes values in a finite alphabet $\m{X}$,
and $B^n$ is uniformly distributed over the set
$\{b^n\in\{0,1\}^n\,:\,n_1(b^n) = m\}$. Then for any $\tau>0$
\begin{equation}\label{eq:hoeffding_sampling}
\Prob\left(\,\left\|\vect{p}_{\rm emp}(Z^n)-\vect{p}_{\rm
emp}\left(Z^n\samp B^n\right)\right\|_\infty>\tau\right) \;\leq\;
2|\m{X}|\exp\left(-2m\tau^2\right)
\end{equation}
\end{lemma}
\begin{proof}
See Appendix \ref{app:lemmas}.
\end{proof}

The following Lemma is an analogue of Lemma
\ref{lem:hoeffding_sampling} for causally independent sampling,
and is a direct consequence of the Azuma-Hoeffding inequality for
bounded-difference martingales \cite{Chung2006}.
\begin{lemma}[Causally independent sampling]\label{lem:azuma_sampling} Let $Z^n,B^n$ be a pair of r.v.
sequences , where $Z^n$ takes values in a finite alphabet $\m{X}$.
Suppose $B_{k+1}\sim{\rm Ber}(q)$ and is statistically independent
of $(B^k,Z^{k+1})$ for any $k\in \brk{n}$. Then for any $\tau>0$,
\begin{align}\label{eq:azuma_sampling}
\Prob\left(\,\left\|\vect{p}_{\rm
emp}(Z^n)-\alpha(B^n)\vect{p}_{\rm emp}\left(Z^n\samp
B^n\right)\right\|_\infty>\tau\right) \;\leq\;
2|\m{X}|\exp\left(-\frac{n\tau^2q^2}{2}\right)
\end{align}
where
\begin{equation}\label{eq:norm_factor}
\alpha(B^n) \;\dfn\; \frac{n_1(B^n)}{\Expt\left\{n_1(B^n)\right\}}
\end{equation}
\end{lemma}
\begin{proof}
See Appendix \ref{app:lemmas}.
\end{proof}
In the context of Lemma \ref{lem:azuma_sampling} above, $B^n$ is
said to be a (i.i.d.) \textit{causal sampling
sequence}\footnote{The i.i.d. sequence $B^n$ is only causally
independent of $Z^n$, but the two sequences may generally be
dependent. For instance, setting $Z_1$ constant and $Z_k =
B_{k-1}$ satisfies the conditions of the Lemma.} for $Z^n$. The
multiplication of the sample's empirical distribution by the
factor $\alpha(\cdot)$ is referred to as
\textit{$\alpha$-normalization}. Note that
$\alpha(B^n)\vect{p}_{\rm emp}\left(Z^n\samp B^n\right)$ is in
fact the vector of symbol occurrences in $Z^n\samp B^n$,
normalized by $\Expt(n_1(B^n))=nq$ instead of by $n_1(B^n)$.
Moreover, $\alpha(B^n)\rightarrow 1$ almost surely (a.s.), hence
this vector converges a.s. to a probability distribution as
$n\rightarrow \infty$.

\section{Channel Model and Main Result}\label{sec:result}

A (causal) \textit{channel} over a common input and output finite
alphabet $\m{X}$, is a sequence of conditional probability
distributions
$\m{W}=\left\{W_k(\cdot|x^k,y^{k-1})\right\}_{k=1}^\infty$ over
$\m{X}$, where $x^k\in\m{X}^k,y^{k-1}\in\m{X}^{k-1}$. Two
sequences of r.v's $(X^\infty,Y^\infty)$ taking values in $\m{X}$
are said to be a pair of \textit{input/output sequences} for the
channel respectively, if for any
$k\in\NaturalF\,,x^k\in\m{X}^k\,,y^k\in\m{X}^k$,
\begin{equation*}
\Prob_{Y_k|X^kY^{k-1}}(y_k|x^k,y^{k-1}) = W_k(y_k|x^k,y^{k-1})
\end{equation*}
We will find it convenient to \textit{model} the channel's action
on its input as the modulo-addition of a \textit{realized noise
sequence} $Z^\infty$ corresponding to $(X^\infty,Y^\infty)$,
implicitly defined by
\begin{equation}\label{eq:realized_noise_def}
Y_k = X_k + Z_k\,,\qquad k\in\NaturalF
\end{equation}
In fact, a channel can be equivalently defined by the conditional
distribution of the noise sequence given past and present inputs
and past outputs $\Prob_{Z_k|X^kY^{k-1}}(z_k|x^k,y^{k-1})$. A
channel $\m{W}$ is called \textit{modulo-additive} if the
following Markov relation is satisfied for any pair of
input/output sequences and any $k\in\NaturalF$,
\begin{equation}\label{eq:Markov_relation_mod_add}
Z_k\leftrightarrow X^{k-1}Y^{k-1} \leftrightarrow X_k
\end{equation}
or equivalently, if $W_k(y_k+z|x_k+z,x^{k-1},y^{k-1})$ is
independent of $z\in\m{X}$ for any $x^k,y^k\in\m{X}^k$. Note that
this definition of a modulo-additive channel allows the noise
sequence to depend on previous inputs and outputs in a general
way. The more restricted class of modulo-additive channels where
the channel is completely defined by the noise distribution
itself, is discussed below.

The family of all causal channels over $\m{X}$ is denoted by
$\mathscr{C}_{\sstm{X}}$, and the family of all modulo-additive
channels over $\m{X}$ is denoted by $\mathscr{M}_{\sstm{X}}\subset
\mathscr{C}_{\sstm{X}}$. The families $\mathscr{C}_{\m{\sstm{X}}}$
and $\mathscr{M}_{\sstm{X}}$ are broad, including also
non-stationary and non-ergodic channels. In particular,
$\mathscr{C}_{\sstm{X}}$ includes the following families of
channels sometimes used for modelling channel uncertainty
\cite{reliable_comm_review}:
\begin{itemize}
\item The \textit{Compound Memoryless Channel}, which is a family
of time-invariant memoryless channels, or in our notation all
channels for which (informally) $W_k(\cdot|x^k,y^{k-1}) =
W(\cdot|x_k)$, where $W(\cdot|\cdot)\in S$ for some set $S$ of
conditional probability distributions over $\m{X}$. Specifically,
$\mathscr{M}_{\sstm{X}}$ includes compound channels for which $S$
consists only of (memoryless) modulo-additive mappings.


\item The \textit{Arbitrarily Varying Channel (AVC)}, which is a
family of time-varying memoryless channels, or in our notation all
channels for which (informally) $W_k(\cdot|x^k,y^{k-1}) =
W_k(\cdot|x_k)$, where each $W_k(\cdot|\cdot)\in S$ for some set
$S$ (state space) of conditional probability distributions over
$\m{X}$. Alternatively, an AVC can also be defined via the noise
sequence by requiring $Z_k\leftrightarrow X_k \leftrightarrow
X^{k-1}Y^{k-1}$. Once again, $\mathscr{M}_{\sstm{X}}$ includes all
AVC's for which $S$ consists of only (memoryless) modulo-additive
mappings.

\item \textit{Noise Sequence Channels}: This family is denoted by
$\mathscr{N}_{\sstm{X}}$, and consists of all (modulo-additive)
channels that are completely defined by the noise sequence itself,
i.e., for which the (stricter) Markov relation
\begin{equation}\label{eq:Markov_relation_noise_seq}
Z_k\leftrightarrow Z^{k-1} \leftrightarrow X^kY^{k-1}
\end{equation}
holds for any $k\in\NaturalF$. Note that some texts use noise
sequence channels as the standard definition for a modulo-additive
channel. Our definition for a modulo-additive channel is broader,
allowing a general coupling between the noise sequence and
previous inputs/outputs, and so $\mathscr{N}_{\sstm{X}}\subset
\mathscr{M}_{\sstm{X}}$ with the inclusion being strict.

\item \textit{Individual Noise Sequence Channels}: This family
consists of all noise sequence channels for which the noise
sequence is an individual sequence $Z^\infty = z^\infty$, i.e.,
$W_k(y_k|x^k,y^{k-1})=\delta_{y_k,x_k+z_k}$ where $\delta_{i,j}$
is Kronecker's delta. It is a subfamily of
$\mathscr{N}_{\sstm{X}}$ defined above, and may be viewed as an
AVC with $S$ being the set of all deterministic modulo-additive
mappings. The example of the binary channel given in the
introduction falls into this category.

\item \textit{Causal Adversarial Channels}: Loosely speaking, a
causal adversarial channel is one for which at each time point an
adversary (jammer) chooses a (possibly random) input-output
mapping according to some (possibly random) \textit{strategy},
that may arbitrarily depend on previous channel inputs and
outputs. It is easy to see that the family of causal adversarial
channels is in fact equivalent to $\mathscr{C}_{\m{\sstm{X}}}$,
since any strategy employed by the adversary can be equivalently
described by the sequence
$\left\{W_k(\cdot|x^k,y^{k-1})\right\}_{k=1}^\infty$. If the
adversary is limited to use only modulo-additive mapping
strategies, then this is equivalent to the family
$\mathscr{M}_{\sstm{X}}$. In the sequel, we will sometimes find it
convenient to use the adversarial point of view.
\end{itemize}

\begin{figure}[htbp]
\leavevmode
\begin{center}
\includegraphics[scale=0.6,angle=0]{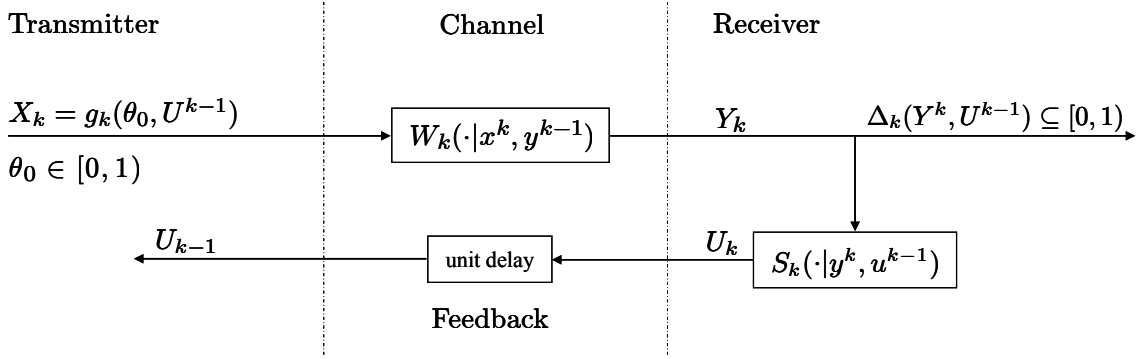}
\caption{Channel model and a feedback transmission
scheme\label{fig:chan_mod}}
\end{center}
\end{figure}

The communications problem with feedback over a channel
$\m{W}\in\mathscr{C}_{\sstm{X}}$ is now described. Without loss of
generality, we assume a \textit{transmitter} is in possession of a
\textit{message point} $\theta_0\in [0,1)$, its binary expansion
representing an infinite bit string to be reliably conveyed to a
\textit{receiver} over the channel $\m{W}$ (later assumed to be
unknown). A (sequential) \textit{feedback transmission scheme} is
described by a triplet $(G,\m{S},\Delta)$, where
$G=\{g_k:[0,1)\times \m{X}^{k-1}\mapsto\m{X}\}_{k=1}^\infty$ is a
sequence of \textit{transmission functions},
$\m{S}\in\mathscr{C}_{\sstm{X}}$ is a \textit{feedback strategy},
$\Delta=\{\Delta_k:\m{X}^k\times\m{X}^{k-1}\mapsto\mathfrak{J}\}_{k=1}^\infty$
is a sequence of \textit{decoding rules}, and $\mathfrak{J}$ is
the set of all binary subintervals of the unit
interval\footnote{There is a one-to-one correspondence between any
finite binary string $b_1b_2\ldots b_k$ and a \textit{binary
subinterval} $[\alpha,\beta)\subseteq[0,1)$ where
$\alpha=0.b_1b_2\ldots b_k$ and $\beta=\alpha+2^{-k}$ (similar to
arithmetic coding).}. A scheme is said to use \textit{passive
feedback} if $\m{S}$ consists of only deterministic conditional
distributions, and is otherwise said to use \textit{active
feedback}. A scheme is said to use \textit{asymptotically passive
feedback} if the portion of non-deterministic conditional
distributions within the first $n$ elements of $\m{S}$ tends to
zero with $n$.

A feedback transmission scheme $(G,\m{S},\Delta)$ used over the
channel $\m{W}\in\mathscr{C}_{\sstm{X}}$ with a message point
$\theta_0\in[0,1)$ is described by the following construction,
also depicted in Figure \ref{fig:chan_mod}:
\begin{itemize}
\item $(X^\infty,Y^\infty)$ is an input/output pair for the
channel $\m{W}$. The channel input sequence is said to be
\textit{generated} by the transmitter, while the channel output
sequence is said to be \textit{observed} by the receiver.

\item $(Y^\infty,U^{\infty})$ is an input/output pair for the
feedback strategy $\m{S}$.

\item The channel input sequence is generated by the transmitter
for any $k\in\NaturalF$, as follows:
\begin{equation}\label{eq:input_gen}
X_k=g_k(\theta_0,U^{k-1})
\end{equation}
The existence of an instantaneous noiseless feedback link is
manifested through the fact that the \textit{feedback sequence}
$U^\infty$, which is causally generated from $Y^\infty$ by the
receiver via the feedback strategy $\m{S}$, is causally available
to the transmitter. Note that passive feedback means that $U_k$ is
a deterministic function of $Y^k$, with the most common example
being when the channel output is fed back to the transmitter,
i.e., $U_k=Y_k$.

\item The following Markov relation is satisfied for any
$k\in\NaturalF$:
\begin{align}\label{eq:markov_relation}
Y_k&\leftrightarrow X^kY^{k-1}\leftrightarrow U^{k-1}
\end{align}
Loosely speaking, this relation guarantees that any randomness
generated by the receiver (and shared with the transmitter via
feedback) is ``private'', i.e., the channel/adversary has no
direct access to it and its actions are based on observing channel
inputs/outputs only.

\item $\Delta_k(Y^k,U^{k-1})$ is the receiver's \textit{decoded
interval} at time $k$.

\end{itemize}
The construction above uniquely determines the joint distribution
of $(X^\infty,Y^\infty,U^\infty)$. If transmission is terminated
at time $n$, the receiver decodes bits that correspond to the
decoded interval $\Delta_n(Y^n,U^{n-1})$ as being the leading bits
in the message point's binary expansion. In accordance, the
associated \textit{rate} and (pointwise) \textit{error
probability} at time $n$ are defined as
\begin{align}\label{eq:rate_err_def}
R_n(\m{W},\theta_0) \dfn
-\frac{1}{n}\,\log\left|\Delta_n(Y^n,U^{n-1})\right|\,,\qquad
p_e(n,\m{W},\theta_0) \dfn
\Prob(\theta_0\not\in\Delta_n(Y^n,U^{n-1}))
\end{align}
Modelling the channel actions as the modulo-addition of a realized
noise sequence $Z^\infty$ as in (\ref{eq:realized_noise_def})
allows us to define the (modulo-additive, zero-order)
\textit{empirical capacity} at time $n$ as
\begin{align}\label{eq:def_emp_cap}
C^{\,\rm emp}_n(\m{W},\theta_0) \dfn \log{|\m{X}|}-H_{\rm
emp}(Z^n)
\end{align}
where the r.v. $H_{\rm emp}(Z^n)$ is the zero-order empirical
entropy of $Z^n$. Hence, the empirical capacity is the capacity of
a corresponding memoryless modulo-additive channel, with a
marginal noise distribution that coincides with the empirical
distribution of $Z^n$. In general, both the instantaneous rate and
the empirical capacity are r.v's with distributions that depend on
the channel, the message point and even the transmission scheme
itself (the latter dependency is suppressed). Note however that in
the special case where communications take place over a noise
sequence channel $\m{W}\in\mathscr{N}_{\sstm{X}}$, the empirical
capacity depends only on the channel, and for an individual noise
sequence channel, it is deterministic.

The universal communication problem over a family of channels
$\mathscr{F}\subseteq\mathscr{C}_{\sstm{X}}$ is now described.
Suppose a feedback transmission scheme $(G,\m{S},\Delta)$ is used
for communication over an \underline{unknown} channel
$\m{W}\in\mathscr{F}$. Regarding the empirical capacity as a
measure for how well the channel behaves, a desirable property
would be for the scheme, although being fixed and independent of
the actual channel in use, to achieve rates close to the empirical
capacity with a low error probability. Making this notion precise,
a scheme $(G,\m{S},\Delta)$ is said to (uniformly) \textit{achieve
the empirical capacity} over the family $\mathscr{F}$, if
\begin{align}\label{eq:achv_emp_cap_def}
\nonumber
&\sup_{\m{W}\in\mathscr{F},\theta_0\in[0,1)}p_e(n,\m{W},\theta_0)
< \eps_1(n)
\\
\inf_{\m{W}\in\mathscr{F},\theta_0\in[0,1)}\Prob\Big{(}&R_n(\m{W},\theta_0)>C^{\,\rm
emp}_n(\m{W},\theta_0)-\eps_2(n)\Big{)}
> 1-\eps_3(n)
\end{align}
where all $\eps_1(n),\eps_2(n),\eps_3(n)\rightarrow 0$. Such a
scheme is also called \textit{universal} for the family $\m{F}$.

In the discussion so far we have considered \textit{horizon-free}
transmission schemes, namely schemes that do not depend on any
decoding deadline and can be terminated at any time. In the
sequel, we also consider \textit{finite-horizon} schemes, which
are schemes that must terminate at some given time $n$ (horizon).
The horizon-free construction and the subsequent definitions of
rate and error probability immediately carry over to the
finite-horizon setting, via simple truncation. A sequence
$\left\{(G,\m{S},\Delta)_n\right\}_{n=1}^\infty$ of finite-horizon
transmission schemes, with $(G,\m{S},\Delta)_n$ having a horizon
$n$, is said to achieve the empirical capacity over a family
$\m{F}$ if for any $n\in\NaturalF$ the scheme $(G,\m{S},\Delta)_n$
satisfies (\ref{eq:achv_emp_cap_def}), and
$\eps_1(n),\eps_2(n),\eps_3(n)\rightarrow 0$. A finite-horizon
scheme is loosely said to achieve the empirical capacity, whenever
it is clear that a suitable sequence of such schemes with an
arbitrarily large horizon can be constructed. We now state our
main result:

\begin{theorem}\label{thrm}
There exists a horizon-free feedback transmission scheme
$(G,\m{S},\Delta)$ using asymptotically passive feedback, that
achieves the empirical capacity over the family
$\mathscr{M}_{\sstm{X}}$. Such a universal scheme is constructed
explicitly below. Furthermore, the scheme can be adapted to
achieve the empirical capacity over the larger family
$\mathscr{C}_{\sstm{X}}$, if common randomness is available.
\end{theorem}
\begin{proof}
The rest of the paper is dedicated to the construction of the
universal scheme and hence to the proof of the Theorem. The
discussion in the body of the paper focuses on a finite-horizon
feedback transmission scheme, which is introduced in section
\ref{sec:scheme}, and shown to be universal for the family
$\mathscr{M}_{\sstm{X}}$ of modulo-additive channels in section
\ref{sec:analysis}. The horizon-free variant of this scheme is
discussed in appendix \ref{app:horizon_free}, and the adaptations
(via common randomness) required to obtain universality for the
family $\mathscr{C}_{\sstm{X}}$ of all causal channels , are
relegated to Appendix \ref{app:dithering}.
\end{proof}
The following remarks are now in order:

\begin{enumerate}[1)]

\item The probabilities in (\ref{eq:rate_err_def}) and
(\ref{eq:achv_emp_cap_def}) are taken over the randomness created
both by the feedback strategy, and by the channel. The randomness
due to feedback is negligible yet essential as manifested by the
special case of an individual noise sequence channel, where
without randomness one is limited by Berlekamp's results
\cite{Berlekamp-Thesis} and the empirical capacity cannot be
attained, even being known in advance. Note also that the
definition in (\ref{eq:achv_emp_cap_def}) requires uniform
convergence over the message point. This is the variable-rate
counterpart of a maximum error probability criterion, and from an
adversarial viewpoint is equivalent to the assumption that the
adversary knows the message point.

\item As already mentioned, when communicating over an unknown
member of $\mathscr{M}_{\sstm{X}}$ or $\mathscr{C}_{\sstm{X}}$ no
rate can be guaranteed in advance since both families include
(many) channels with zero capacity. Furthermore, since the channel
law may vary arbitrarily there is no hope to identify the actual
channel in use and attain its capacity, even with feedback. Our
approach is more ``optimistic'': We disregard the complex nature
of $\mathscr{M}_{\sstm{X}}$, $\mathscr{C}_{\sstm{X}}$ and model
the channel actions as being memoryless modulo-additive, although
these are usually not. This simple model allows us to
opportunistically attain rates that correspond to the empirical
goodness of the realized channel (measured relative to our model),
no matter what the true channel law is. In the special case of
noise sequence channels, we obtain universality w.r.t. any
competing scheme that is informed of the empirical distribution of
the noise sequence in advance. Of course, more complex models for
channel actions can be considered. For instance, one can model the
actions as being modulo-additive with some Markovian statistical
dependence, or as being memoryless but input dependent. These more
elaborate models allow to universally approach suitably defined
(and possibly higher) empirical capacities, and are discussed in
part II of this work \cite{empirical_capacity_partII}.

\item The empirical capacity over the family
$\mathscr{M}_{\sstm{X}}$ is achieved essentially without common
randomness, since the negligible amount nevertheless required can
be generated via feedback. However, when communicating over the
family $\mathscr{C}_{\sstm{X}}$ we require common randomness that
cannot be accommodated by feedback. As described in Appendix
\ref{app:dithering}, this randomness is chiefly used for
dithering, i.e., making the input distribution uniform. This is
not merely an artifact, but has to do with the fact that the
empirical capacity is defined in terms of a noise sequence, and
for channels in $\mathscr{C}_{\sstm{X}}$ the empirical
distribution of the realized noise sequence depends on the
empirical input distribution, a dependence which a modulo-additive
model cannot capture. For instance, consider the extreme case of a
binary channel where the channel's output at each time point is
randomly chosen in an i.i.d. fashion to be $\sim{\rm Ber}(\eps)$,
independently of the inputs. This memoryless channel is in
$\mathscr{C}_{\{0,1\hspace{-0.04cm}\}}$ (but not in
$\mathscr{M}_{\{0,1\hspace{-0.04cm}\}}$), and its capacity is of
course zero. Suppose one tries to communicate over this channel
using some transmission scheme, and at the end of transmission the
empirical distribution of the inputs turns out to be $q$. Then
with high probability, the empirical distribution of the realized
noise sequence will be close to $q*\eps\dfn q(1-\eps)+(1-q)\eps$,
and the empirical capacity will therefore be close to $1-h_{\sst
B}(q*\eps)$ which is positive for $q\neq \frac{1}{2}$. This
example demonstrates that for the family $\mathscr{C}_{\sstm{X}}$,
unless the input distribution is guaranteed to be close to uniform
with high probability, the empirical capacity as defined may not
be the right quantity to look at.

\item When $\m{W}\in\mathscr{M}_{\sstm{X}}$ happens to be a
memoryless channel, the empirical capacity converges a.s. to the
classical capacity of the channel. This is of course generally
untrue for memoryless channels $\m{W}\in\mathscr{C}_{\sstm{X}}$
(with dithering). It is straightforward that due to the
modulo-additive modelling, the empirical capacity cannot exceed
the mutual information of $\m{W}$ with a uniform input, but in
fact the penalty may even be larger. For example, consider a
general binary memoryless channel
$\m{W}\in\mathscr{C}_{\{0,1\hspace{-0.04cm}\}}$, described by
\begin{equation*}
W_k(j\,|\,x_k=i,x^{k-1},y^{k-1}) = p_{ij} \qquad i,j\in\{0,1\}
\end{equation*}
With a uniform input (obtained via dithering), the empirical
capacity will converge a.s. to $1-h_{\sst
B}(\frac{1}{2}(p_{00}+p_{11}))$. This quantity is the capacity of
a BSC obtained by averaging the channel $\m{W}$ with its
``cyclicly shifted'' counterpart, which is a binary memoryless
channel characterized by the transition probabilities
$q_{ij}=p_{i+1,j+1}$ (modulo addition). By the convexity of the
mutual information in the transition matrix, and due to the
symmetry between $\m{W}$ and its cyclically shifted counterpart,
the capacity of this BSC is upper bounded by the mutual
information of $\m{W}$ with a uniform input. Furthermore, unless
$\m{W}$ happens to be a BSC to begin with, this inequality is
strict and the empirical capacity is a.s. strictly smaller than
the mutual information of $\m{W}$ with a uniform input. The
discussion is easily extended to larger alphabets. This point and
related issues mentioned in the previous remark, are further
pursued in part II of this work \cite{empirical_capacity_partII}.

\end{enumerate}

\section{The Universal Scheme}\label{sec:scheme}

In this section we introduce a finite-horizon transmission scheme
achieving the empirical capacity over the family
$\mathscr{M}_{\m{X}}$. We find it instructive to focus our
discussion on this setting, as it is simpler yet includes all the
core ideas. The more exhaustive horizon-free scheme and its
extension to the larger family $\m{C}_{\sstm{X}}$ using common
randomness, are discussed in Appendices \ref{app:horizon_free} and
\ref{app:dithering}. We start by building intuition for the binary
alphabet case, followed by a step by step construction of a binary
alphabet universal scheme. This scheme is then generalized to a
finite alphabet setting via some simple modifications. The rate
and error probability analysis of the scheme appears in section
\ref{sec:analysis}. In this section, transmission is assumed to
take place over a fixed period of $n$ channel uses.

\subsection{The Horstein Scheme for the
BSC}\label{subsec:horstein} We first discuss the simple case where
the channel in use is known to be a BSC with a given crossover
probability $p$, which in our terminology means a noise sequence
channel (i.e., one satisfying the Markov relation
(\ref{eq:Markov_relation_noise_seq})) with an i.i.d. $\sim{\rm
Ber}(p)$ noise sequence $Z^\infty$. For this setting, we describe
the well known passive feedback transmission scheme proposed by
Horstein \cite{horstein}. In that scheme, the message point is
assumed to be selected at random uniformly over the unit interval.
The receiver constantly calculates the a-posteriori probability
distribution of the message point given the bits it has seen so
far. These bits are passively fed back to the transmitter (namely
$U_k=Y_k$ in our terminology), which can therefore calculate the
posterior as well. A zero or one is transmitted according to
whether the message point currently lies to the left or to the
right of the posterior's median point. Thus the transmitter always
answers the most informative yes/no question that can be posed by
the receiver.

Specifically, let $\Theta_0$ be the random message point and
denote its posterior density at time $k$ (given the observed
outputs) by $f_k(\theta)\dfn f_{\Theta_0|Y^k}(\theta|y^k)$ for
$\theta\in[0,1)$. Denote the median point corresponding to
$f_k(\theta)$ by $\mu_k$. Since $\Theta_0$ is uniform over the
unit interval, we have $f_0(\theta) = \ind_{[0,1)}(\theta)$ and
$\mu_0=\frac{1}{2}$. The transmission functions are hence given by
\begin{equation*}
g_k\left(\theta_0,y^{k-1}\right) = \left \{ \begin{array} {lc} 0 &
\theta_0 < \mu_{k-1} \\ 1 & \theta_0
> \mu_{k-1}
\end{array} \right .
\end{equation*}
and the transition from $f_k(\theta)$ to $f_{k+1}(\theta)$ is
given by:
\begin{equation*}
f_{k+1}(\theta) = \left \{ \begin{array} {lc} 2(p y_{k+1} + q
(1-y_{k+1}))f_k(\theta) & \theta < \mu_k \\ 2(q y_{k+1} + p
(1-y_{k+1}))f_k(\theta) & \theta > \mu_k
\end{array} \right .
\end{equation*}
where $q=1-p\,$. The transition from $f_k(\theta)$ to
$f_{k+1}(\theta)$ and the corresponding transmission of
$X_k=g_k\left(\theta_0,Y^{k-1}\right)$ are referred to in the
sequel as a \textit{Horstein iteration}. Several optimal decoding
rules are associated with the Horstein scheme. A fixed rate $R$
rule is to decode the binary interval of size $2^{-\lfloor
nR\rfloor}$ with the maximal posterior probability, which is our
notations reads
\begin{equation*}
\Delta_n(y^n,u^{n-1})=\Delta_n(y^n) =
\argsup_{I\in\mathfrak{J}\,,|I|=2^{-\lfloor
nR\rfloor}}\int_{I}f_{\sst\Theta_0|Y^n}(\theta|y^n)d\theta
\end{equation*}
A variable rate rule with a target error probability $p_e$, is to
decode the smallest binary interval with a posterior probability
exceeding a threshold $1-p_e$. There is also the bit-level
decoding rule in which a bit is decoded whenever its corresponding
binary interval has accumulated a posterior probability greater
than $1-p_e$, where $p_e$ is a target probability of bit
error\footnote{For instance, when the posterior probability
(w.r.t. $f_k(\theta)$) of either $\left[0,\frac{1}{2}\right)$ or
$\left[\frac{1}{2},1\right)$ exceeds $1-p_e$, the MSB of the
message point is decoded as either 0 or 1 respectively.}. The
Horstein scheme has been long conjectured to achieve the capacity
of the BSC with either a fixed or a variable rate decoding rule,
but this fact was proved in rigor only recently
\cite{posterior_matching_isit08}.

\subsection{Binary Channels with Noise Constraints}
Let us now take a step towards the unknown channel setting by
considering a subfamily of $\mathscr{M}_{\{0,1\hspace{-0.04cm}\}}$
where the empirical distribution of the noise sequence is known in
advance to a.s. satisfy $p_{\rm emp}(Z^n)<p<\frac{1}{2}$ (e.g., an
individual noise sequence with a fractions of `1's smaller than
$p$). From an adversarial point of view, this can be thought of as
imposing a power constraint on the adversary. A plausible idea
would be to communicate by performing Horstein iterations using
$p$ in lieu of the crossover probability, hoping that the average
performance of the scheme in the BSC setting will carry over to
this more stringent setting, i.e., enable to achieve $1-h_{\sst
B}(p)$ uniformly over the noise-constrained family. Unfortunately,
this is not the case since Berlekamp's results
\cite{Berlekamp-Thesis} imply that for many values of $p$ there
exist pairs of message points and individual noise sequences
(satisfying the constraint) for which decoding will surely fail.
Nevertheless, as we now show, there is little information missing
at the receiver to get it right.

We now make two key observations regarding the Horstein
transmission process. First, we notice that $f_k(\theta)$ is a
quasi-constant function, over at most $k+1$ distinct intervals
whose union is the unit interval. Second, when transmission is
terminated after $n$ channel uses, we have
\begin{equation}\label{eq:fn1}
f_n(\theta=\theta_0) = 2^n (1-p)^{n_0(Z^n)}p^{n_1(Z^n)} = 2^n
(1-p)^{n(1-p_{\rm emp}(Z^n))}p^{np_{\rm emp}(Z^n)}
\end{equation}
where $\theta_0$ is the message point. This stems directly from
the fact that $\theta_0$ is always on the correct side of the
median, so that its density is multiplied by $2(1-p)$ when there
is no error, and by $2p$ otherwise. Now, let the \textit{message
interval} at time $k$ be the interval containing $\theta_0$ over
which $f_k(\theta)$ is constant, and let $2^{-\ell}$ be its length
at the end of transmission, for some $\ell>0$. Using
(\ref{eq:fn1}) we have that
\begin{equation*}
2^{-\ell}\cdot f_n(\theta=\theta_0) \leq 1 \quad \Rightarrow \quad
\ell \geq n\big{(}1-h_{\sst B}(p_{\rm
emp}(Z^n)\big{)}-D\big{(}p_{\rm emp}(Z^n) \,\|\, p\big{)}
\end{equation*}
Now, assume the decoder could identify with certainty which is the
message interval at the end of transmission. In that case, the
common most significant bits in the binary expansion of points
inside the message interval (which also correspond to the message
point itself) could be decoded, error free! This means that an
instantaneous decoding rate of
\begin{equation}\label{eq:rate1}
R_n = \frac{\lfloor\ell\rfloor}{n} \geq 1-h_{\sst B}(p_{\rm
emp}(Z^n))-D\big{(}p_{\rm emp}(Z^n) \,\|\, p\big{)} - \frac{1}{n}
\end{equation}
could be attained. Actually, another information bit is required
to allow the above rate, as the message interval may sometimes be
inconveniently located over the binary grid and not enough bits
(if any) can be decoded. We further elaborate on this point in
subsection \ref{subsec:full_scheme} when the full scheme is
presented. Notice that the expression in (\ref{eq:rate1}) can be
divided into two parts: $1-h_{\sst B}(p_{\rm emp})$ is the
empirical capacity of the channel, and $D(p_{\rm emp} \,\|\, p)$
is a penalty term for using the maximal value $p$ of $p_{\rm emp}$
instead of $p_{\rm emp}$ itself. Note also that
\begin{equation}\label{eq:inf_rate}
\inf_{p_{\rm emp}<p}R_n \;\geq\; 1-h_{\sst B}(p) - \frac{1}{n}
\end{equation}
and therefore the rate attained by this variable rate scheme (had
the message interval been known at the end of transmission) is
guaranteed to be asymptotically no less than $1-h_{\sst B}(p)$,
for \textit{any} message point $\theta_0\in[0,1)$.

As observed before, there are at most $n+1$ distinct intervals
over which $f_n(\theta)$ is constant, therefore no more than
$\lceil\log{(n+1)}\rceil$ bits of side information are required in
order to identify the message interval at the end of transmission.
This means that the decoding rate in (\ref{eq:rate1}) is
achievable (error free) if only $1+\lceil\log{(n+1)}\rceil$ bits
could be reliably conveyed to the receiver at the end of
transmission. Thus, while \cite{Berlekamp-Thesis} determines that
it is generally impossible to communicate at such rate with no
errors, the size of the decoding uncertainty preventing us from
attaining it is very small. For instance, if list decoding is
allowed then (\ref{eq:inf_rate}) could be attained using a list
whose size grows only linearly (and not exponentially) with $n$.

In the following subsections we present a simple randomization
technique by which these extra bits can be reliably conveyed to
the receiver, with no asymptotic decrease in the data rate, so
that the decoder can determine with high probability the correct
message from the list. This technique requires a sub-linear number
of random bits shared by the transmitter and the receiver
(obtained via feedback, or possibly via a common random source).
Moreover, through a sequential use of randomness we will be able
to present a feasible transmission scheme that tracks the
empirical distribution of the realized noise sequence, so that a
significantly higher rate approaching the \textit{empirical
capacity} $1-h_{\sst B}(p_{\rm emp}(Z^n))$ is attained, avoiding
the penalty term in (\ref{eq:rate1}).

\subsection{Sequential Probability Estimation}\label{subsec:seq_prob_est}
We now turn to the general case where the channel in use is an
arbitrary unknown member of
$\mathscr{M}_{\{0,1\hspace{-0.04cm}\}}$, i.e., from the
adversarial point of view there are no constraints (besides
causality) on the way the noise sequence is generated by the
adversary (e.g., the noise may be some unknown individual
sequence). A reasonable idea could be to plug in a sequential
estimator for the empirical distribution of the noise $p_{\rm
emp}(Z^k)$ into the Horstein iterations, that is, to let the
``crossover probability'' used by the receiver to calculate the so
called ``posteriori distribution'' of the message point, vary with
time. Specifically, this amounts to
\begin{equation*}
f_{k+1}(\theta) = \left \{ \begin{array} {lc}
2\big{(}\widehat{p}_{k+1}(Z^k)
y_{k+1} + \widehat{q}_{k+1}(Z^k) (1-y_{k+1})\big{)}f_k(\theta) & \theta < \mu_k \\
2\big{(}\widehat{q}_{k+1}(Z^k) y_{k+1} + \widehat{p}_{k+1}(Z^k)
(1-y_{k+1})\big{)}f_k(\theta) & \theta > \mu_k
\end{array} \right .
\end{equation*}
where $\widehat{p}_{k+1}$ is a sequential estimator applied to the
noise sequence $Z^k$ and $\widehat{q}_{k+1}\dfn
1-\widehat{p}_{k+1}$. Note that $f_k(\theta)$ is still a
probability density function, but looses the meaning of a true
posterior in this unknown channel setting. We henceforth loosely
refer to $f_k(\theta)$ as the \textit{empirical posterior} of the
message point (relative to the estimator in use).

This idea is of course problematic, since the noise sequence is
causally known only to the transmitter and not to the receiver,
but for the moment let us assume that the estimates can be somehow
made known to the receiver, and take care of this point later. The
first of the two key observations from the previous subsection
still holds, i.e., $f_k(\theta)$ is quasi-constant over $k+1$
disjoint intervals whose union is the unit interval. The empirical
posterior evaluated at the message point at the end of
transmission is now equal to
\begin{equation*}
f_n(\theta=\theta_0) = 2^n \prod_{k=1}^n
\widehat{p}_k^{\;Z_k}(1-\widehat{p}_k)^{1-Z_k} =
2^n\widehat{p}\,(Z^n)
\end{equation*}
where $\widehat{p}\,(Z^n)$ is the probability assigned to the
entire noise sequence $Z^n$ by the estimator in use. Using this
fact and assuming again that at the end of transmission we know
which one of the intervals is the message interval (which is then
set to be the decoded interval), the instantaneous decoding rate
attained is given by
\begin{equation*}
R_n \geq \frac{1}{n}\lfloor\log{2^n\widehat{p}\,(Z^n)}\rfloor \geq
1- \frac{1}{n}\log\widehat{p}\,(Z^n) - \frac{1}{n}
\end{equation*}
so the shorter the codelength assigned by the estimator to the
noise sequence, or the more \textit{compressible} the strategy of
the adversary is (w.r.t. a memoryless modulo-additive model), the
higher the achieved rate. It is therefore only reasonable to make
use of the KT estimator, or more generally an intermittently
updated KT($b$) estimator. Applying Lemmas \ref{lem:KT} and
\ref{lem:KTb}, the instantaneous decoding rate achieved when using
a KT($b$) estimator is
\begin{equation}\label{eq:rate2}
R_n \geq 1- \frac{1}{n}\log\widehat{p}^{\,\sst{KT(b)}}\,(Z^n) -
\frac{1}{n} \geq 1-h_{\sst B}(p_{\rm emp}(Z^n)) -
K_1\;\frac{b\log{n}}{n} = C^{\,\rm emp}_n(\m{W},\theta_0) -
K_1\;\frac{b\log{n}}{n}
\end{equation}
where $K_1>0$ is constant. Thus, if $bn^{-1}\log{n}=\lito(1)$ the
empirical capacity is asymptotically attained. This holds however
only under the assumptions that the receiver knows the KT($b$)
estimates online, and can also recognize the message interval with
certainty at the end of transmission.

There are two key elements that make this approach work. First,
the update information required by the receiver so that the
assumptions above are satisfied can be made negligible, namely
have rate zero. The message interval is one of at most $n+1$
possible intervals hence requires only $\lceil\log(n+1)\rceil$
bits, and using a KT($b$) estimator only requires to communicate
the number of '1's in the last $b$ channel uses, which requires
$\lceil\log{b}\rceil$ bits per $b$ channel uses and is negligible
if $b^{-1}\log{b}=\lito(1)$. Note the core tradeoff between a
small $b$ required to obtain a small redundancy term in
(\ref{eq:rate2}), and a large $b$ required to make the update
information rate negligible. Second, as we shall see it is
possible to obtain reliable zero rate communications over an
unknown member of $\mathscr{M}_{\{0,1\hspace{-0.04cm}\}}$ as long
as the empirical capacity is not too small, and that the latter
condition can be identified with high probability. These two
observations allow us to make the seemingly unfeasible approach
described so far into a practical scheme that achieves the
empirical capacity.

\subsection{A Universal Binary Alphabet Scheme}\label{subsec:full_scheme}
In this subsection we introduce the universal scheme achieving the
empirical capacity for the binary alphabet, finite-horizon case.
Let us first provide a rough outline of the scheme. Transmission
takes place over a period of $n$ channel uses, which is divided
into blocks of equal length $b=b(n)$. Inside each block,
\textit{Horstein iterations} are performed over the majority of
channel uses, always using the most updated KT estimate.
\textit{Update information} containing the number of '1's in the
previously accepted block together with the index of the current
message interval, is coded using a repetition code and passed to
the receiver over randomly selected positions inside the block,
which are selected via feedback. The idea is that since positions
are random, the ``effective'' channel for the update information
transmission is roughly a BSC with transition probabilities close
to the empirical distribution of the noise sequence inside the
block. This distribution is estimated using a randomly positioned
training sequence, and if the estimation is too close to being
uniform, the block is discarded. Otherwise, the update information
can be reliably decoded with high probability. Loosely speaking,
the discarding process partitions the noise sequence into a
``good'' part and a ``bad'' part, and with high probability the
empirical capacity of the latter part is small. Therefore,
discarding the ``bad'' part increases the rate with high
probability, due to the concavity of the entropy.

If a block is accepted then the polarity of the ``effective''
crossover probability (i.e., above/below $\frac{1}{2}$) can be
reliably determined, and hence the update information (which has a
negligible rate) can be reliably decoded. Once the update
information is successfully decoded, the receiver uses the number
of '1's in the noise sequence from the previous block to update
the KT estimate, which is then used for communications in the next
block. At the end of transmission, the last known message interval
is set to be the decoded interval.

What takes place inside each block is now described in detail. We
define four types of positions within the block - \textit{regular
positions} over which Horstein iterations are performed,
\textit{training positions} over which a training sequence is
transmitted, \textit{update positions} over which update
information is transmitted, and \textit{active feedback positions}
used to select the random positions for the other types. The
non-active feedback positions (regular, training and update) are
referred to as \textit{passive feedback positions}. Apart from
active feedback positions, the receiver passively feeds back what
it receives (i.e., $U_k=Y_k$ over these positions).

\begin{enumerate}[(A)]
\item \label{step1} \textbf{Random positions generation (active
feedback)}: We set a parameter $m = m(n)$ which will indirectly
determine the number of non-regular positions. The active feedback
positions are always at the beginning of the block, and occupy
exactly $b_a=b_a(n)$ positions where $b_a$ is determined in the
sequel as a function of $m,b$. The active positions are used in
order to synchronize the terminals regarding the type of each
passive position that follows. The number of passive positions is
fixed and given by $b_p\dfn b-b_a$. The type of each of the $b_p$
passive positions is determined by an i.i.d. sequence
$\Lambda^{b_p}$ over the alphabet $\{training,update,regular\}$
with a marginal distribution given by
$\left(\frac{m}{b_p},\frac{m}{b_p},1-\frac{2m}{b_p}\right)$. The
selection of $\Lambda^{b_p}$ is synchronized between the
transmitter and the receiver as follows:
\begin{enumerate}[({A}1)]
\item The receiver randomly selects the sequence $\Lambda^{b_p}$
according to the i.i.d. distribution above. Let the r.v's
$(M_t,M_u,M_r)$ denote the number of occurrences of the
corresponding symbols in $\Lambda^{b_p}$, and note that
$\Expt(M_t,M_u,M_r) = (m,m,b_p-2m)$.

\item The type of the sequence\footnote{Note that by \textit{type}
of a sequence we refer to the vector of symbols occurrences, not
to be confused with position types.} $\Lambda^{b_p}$ is then
binary encoded and sent via feedback over active positions, which
requires no more than $2\left\lceil\log{b}\right\rceil$ bits.

\item \label{item:a3} If both $\frac{m}{2} \leq M_t,M_u\leq 2m$
(not too little or too many training and update positions) then
the index of the sequence $\Lambda^{b_p}$ within its
type\footnotemark[\value{footnote}]  is communicated via the
feedback (active positions), which requires no more than
$4m\lceil\log{b}\rceil$ bits. Otherwise, the transmitter overrides
the receiver's selection, and randomly selects the sequences
$\Lambda^{b_p}$ itself.
\end{enumerate}

Now, another sequence $\Gamma^{M_u}$ is selected, determining
which of the update bits is to be transmitted over which update
position (with repetitions). The number of update bits (see step
(\ref{item:update}) below) is $2\lceil\log(n+1)\rceil$, hence
$\Gamma^{M_u}$ is selected in a uniform i.i.d. fashion over the
alphabet $\brk{2\lceil\log(n+1)\rceil}$. If both $\frac{m}{2} \leq
M_t,M_u\leq 2m$, then $\Gamma^{M_u}$ is selected by the receiver,
binary encoded using no more than $2m\big{\lceil}
1+\log\lceil\log(n+1)\rceil)\big{\rceil}$ bits and sent via
feedback over active positions. Otherwise, $\Gamma^{M_u}$ is
selected by the transmitter.

Let us now assume that $b>\log{(n+1)}$, and set the total number
of active positions to $b_a \dfn 8m\lceil\log{b}\rceil$, which is
sufficient to accommodate the synchronization process described
above. If $b^{-1}m\log{b}=\lito(1)$, then this amount becomes
negligible and the feedback strategy is asymptotically passive.
Figure \ref{fig:block} depicts a ``typical'' position assignment
within a block.

\begin{figure}[ht]
\leavevmode
\begin{center}
\includegraphics[scale=1.1,angle=0]{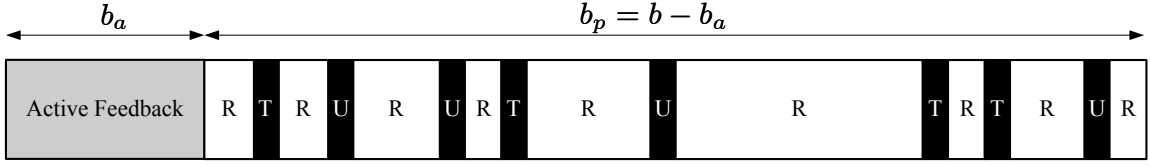}
\caption{A block always begins with $b_a$ active feedback
positions, followed by a bulk of \textbf{R}egular positions that
are randomly replaced with (on the average) $m$ \textbf{T}raining
positions and $m$ \textbf{U}pdate positions.\label{fig:block}}
\end{center}
\end{figure}

\item \textbf{Training transmission}:\label{item:training} A
training sequence is transmitted over the $M_t$ random positions
as determined by $\Lambda^{b_p}$. At the end of the block, the
receiver calculates the \textit{training estimate} $p^{\rm train}$
for the empirical distribution of the noise in the block, which is
a coarse estimate later used for block discarding. Let
$\tilde{Z}^{b_p}$ denote the noise sequence over passive positions
within the current block, i.e., if this is the $k$th block then
$\tilde{Z}^b=Z_{bk-b_p+1}^{bk}$. Let $B^{b_p}$ be the
corresponding \textit{training pattern sequence}, i.e.,
$B_k=\ind_{\rm training}(\Lambda_k)$. The training estimate is set
to
\begin{equation}\label{eq:training_estimate}
\vect{p}^{\rm train} \;\dfn\;
\alpha\left(B^{b_p}\right)\vect{p}_{\rm
emp}\left(\tilde{Z}^{b_p}\samp B^{b_p}\right) =
\left(\frac{M_t}{m}\right)\cdot\vect{p}_{\rm
emp}\left(\tilde{Z}^{b_p}\samp B^{b_p}\right)
\end{equation}
where the $\alpha$-normalization factor is defined in
(\ref{eq:norm_factor}).

\item \textbf{Update transmission}: \label{item:update}Update
information is transmitted over the $M_u$ random positions
determined by $\Lambda^{b_p}$. The uncoded update information
includes the following quantities, all binary encoded:
\begin{enumerate}[(C1)]
\item The number of '1's in the noise sequence over regular
positions in the previously accepted block ($\lceil \log{b}\rceil$
bits).

\item The index of the message interval w.r.t. the interval
partitioning of the empirical posterior at the end of the
previously accepted block ($\lceil\log{(n+1)}\rceil$ bits at the
most).

\item One \textit{ambiguity resolving bit} which is discussed
later on.
\end{enumerate}
The number of uncoded update bits in total is therefore no more
than $2\lceil\log(n+1)\rceil$, and for simplicity we assume that
exactly $2\lceil\log(n+1)\rceil$ uncoded update bits are to be
transmitted (and e.g. zero pad if necessary). Now, on the $k$th
update position (determined by $\Lambda^{b_p}$) the transmitter
sends the $\Gamma_k$-th uncoded update bit. By properly tuning the
scheme parameters we typically have $M_u\gg \log{n}$, hence each
update bit is coded using a repetition code with a random number
of repetitions.

\item \textbf{Horstein iterations with KT($b$) estimates}:
Horstein iterations are performed over the (random) $M_r$ regular
positions as determined by $\Lambda^{b_p}$. The ``crossover
probability'' used is the most updated KT estimate of the
empirical noise distribution available to the receiver. On the
k$th$ block, this estimate is given by
\begin{equation*}
\widehat{p}^{(k)} =
\frac{\frac{1}{2}+\sum_{j=1}^{k-2}\widehat{n}_1(j)I(j)}{1+\sum_{j=1}^{k-2}M_r(j)I(j)}\,,\qquad
\end{equation*}
where $\widehat{n}_1(j)$ is the number of '1's the receiver
assumes appeared in the noise sequence over regular positions in
the $j$th block, as communicated by the update information so far
(may be different than the actual number due to errors), $M_r(j)$
is the value of $M_r$ (number of regular positions) on the $j$th
block, and $I(j)$ is an indicator function that evaluates to one
if the $j$th block was accepted, and to zero if it was discarded.
Note that the estimator works on the sequence of accepted
positions, and is always two accepted blocks behind.

\item \textbf{Block discarding}: The block is discarded if either
$M_t,M_u$ are out of range (in the sense of (\ref{item:a3})), or
if
\begin{equation}\label{eq:discarding_criteria}
\Linf{\vect{p}^{\rm train}-\vect{p}_u} < \tau_{\sst d}
\end{equation}
for some \textit{discarding threshold} $\tau_{\sst d}(n)
=\lito(1)$, where $\vect{p}_u$ is the uniform distribution over
$\{0,1\}$. Otherwise, the block is accepted. When a block is
discarded, the transmitter and receiver return to the state they
were in before the block has started.

\item \textbf{Update information Decoding}: For an accepted block,
the update information is decoded according to the estimated noise
probability, as follows. Let $\tilde{Y}^{b_p}$ denote the channel
output sequence over passive positions within the current block,
i.e., if this is the $k$th block then
$\tilde{Y}^b=Y_{bk-b_p+1}^{bk}$. Let $B^{b_p}_{(i)}$ be the
\textit{repetition pattern sequence} of the $i$th update bit (as
determined by $\Lambda^b,\Gamma^{M_u}$), i.e., a binary sequence
with `1's only in update positions that correspond to a repetition
of that bit. For any $i\in\brk{2\lceil\log{n}\rceil}$, the
receiver calculates the following \textit{update estimate} for the
$i$th bit:
\begin{equation*}
\vect{p}^{{\rm upd},i}\;\dfn\;
\alpha\left(B^{b_p}_{(i)}\right)\vect{p}_{\rm
emp}\left(\tilde{Y}^{b_p}\samp B^{b_p}_{(i)}\right)=
\left(\frac{n_i\left(\Gamma^{M_u}\right)}{m\slash(2\lceil\log(n+1)\rceil)}\right)\cdot\vect{p}_{\rm
emp}\left(\tilde{Y}^{b_p}\samp B^{b_p}_{(i)}\right)
\end{equation*}
where $\alpha$-normalization is used again. Intuitively, we expect
the update estimate to be close to the training estimate only when
the corresponding update bit was a `0', unless the noise sequence
within the block is close to uniform in which case it is likely to
be discarded anyway. Accordingly, the decision rule for the $i$th
update bit is given by
\begin{equation}\label{eq:update_dec_rule}
\Linf{\vect{p}^{{\rm upd},i}-\vect{p}^{\rm train}}
\;\decision{0}{1} \;\;\tau_u
\end{equation}
for some \textit{update decision threshold} $\tau_u(n)=\lito(1)$,
where in case of an equality a `0' is decoded. The decoded
information is used to update the KT estimate, and to store the
new identity of the message interval.

\end{enumerate}

\textbf{Decoding Rule:} Ideally, when transmission ends one would
like to decode the minimal binary interval (i.e., its
corresponding MSB's) containing the last message interval given by
the update information, which (if not error has occurred) contains
the message point. However, as happens in arithmetic coding,
sometimes this minimal binary interval is much larger than the
message interval itself (for instance, if the message interval
contains the point $\frac{1}{2}$ then we cannot decode even a
single bit). To solve this, note that it is possible to divide the
interval $[0,1)$ into binary intervals of size corresponding to
the message interval's size, such that the message interval
intersects no more than two of those, and the only uncertainty
that may be left is which one. The \textit{ambiguity resolving
bit} mentioned in step (\ref{item:update}) above is used to that
end. The receiver thus uses the following decoding
rule\footnote{If all blocks were discarded, the decoded interval
is trivially taken to be $[0,1)$.}: Seek the two smallest adjacent
binary intervals (of the same size) whose union contains the last
known message interval, and decode one of them according to the
last known ambiguity resolving bit. This rule guarantees that the
decoded interval is less than twice the size of the last message
interval.

As we show in section \ref{sec:analysis}, by properly selecting
the dependence of the scheme parameters on $n$, the update
information is guaranteed to be correctly decoded with probability
approaching one without causing any asymptotic decrease in the
data rate, thus allowing the empirical capacity to be approached.

\subsection{A Universal Finite Alphabet Scheme}\label{subsec:modulo-additive}
We now describe the (finite-horizon) finite alphabet $\m{X}$
variant of the universal scheme. Suppose the transmitter and the
receiver can agree on a sequential KT$(b)$ estimator
$\widehat{p}_k^{\;\sst{KT(b)}}(\cdot|Z^{k-1})$ for the noise
sequence at each time $k$ (as in subsection
\ref{subsec:seq_prob_est}). Horstein iterations using this
estimate are performed as follows. The empirical posterior is
initialized as before to a uniform distribution over the unit
interval $f_0(\theta) = \ind_{[0,1)}(\theta)$. At each time point
$k$, unit interval is divided into $|\m{X}|$ consecutive
subintervals with identical probability $|\m{X}|^{-1}$ under the
empirical posterior $f_{k-1}(\theta)$, and the transmitter sends a
symbol that corresponds to the subinterval containing the message
point $\theta_0$. Upon receiving $Y_k\in\m{X}$, the receiver
generates the new empirical posterior $f_{k}(\theta)$ by
multiplying $f_{k-1}(\theta)$ in the interval corresponding to the
symbol $i\in\m{X}$ by the factor
$|\m{X}|\widehat{p}_k^{\;\sst{KT(b)}}(Y_k-i|Z^{k-1})$, where the
minus sign is the modulo-subtraction operator\footnote{The
probability of the $i$th interval under $f_k(\theta)$ is exactly
$\widehat{p}_k^{\;\sst{KT(b)}}(Y_k-i|Z^{k-1})$, hence
$f_k(\theta)$ is a probability distribution.}. Thus the message
point is always in the interval multiplied by the estimate
corresponding to the next value of the noise, and hence we have
\begin{equation*}
f_n(\theta=\theta_0) = |\m{X}|^n\widehat{p}^{\;\sst{KT(b)}}(Z^n)
\end{equation*}
where $\widehat{p}^{\;\sst{KT(b)}}(Z^n)$ is the probability
assigned to the entire noise sequence by the KT$(b)$ estimator.
Similarly to the binary case, there are no more than
$n\left(|\m{X}|-1\right)+1$ subintervals over which $f_n(\theta)$
is constant. Assuming further that the message interval index is
known to the receiver at the end of transmission, then using
Lemmas \ref{lem:KT} and \ref{lem:KTb} for the KT$(b)$ redundancy,
the achieved rate is given by
\begin{equation}\label{eq:rate_pure_fin_alph}
R_n \;\geq \;\log|\m{X}| - H_{\rm emp}(Z^n)
-K_2\frac{|\m{X}|b\log{n}}{n} = C^{\,\rm emp}_n(\m{W},\theta_0)
-K_2\frac{|\m{X}|b\log{n}}{n}
\end{equation}
for some $K_2>0$. Once again, the update information required by
the receiver for the above assumptions to hold incurs in a
vanishing rate penalty, and can be reliably transmitted over
random positions if the empirical capacity is not too small.

The finite alphabet scheme now follows a similar path to that of
the binary alphabet scheme. The transmission period is divided
into blocks of equal size $b=b(n)$, and inside each block we again
have the active feedback, training, update and regular positions.
However, the update information transmission and decoding is
somewhat more involved in this case. We now describe what happens
inside each block:

\begin{enumerate}[(A)]
\item \label{step1X} \textbf{Random position generation (active
feedback)}: We use the same parameters $b,m,b_a$, generate the
corresponding r.v.'s $M_t,M_u,M_r$, and pick the sequence
$\Lambda^{b_p}$ the same way\footnote{Note it is possible to
reduce the number of active positions $b_a$ since the feedback has
a larger capacity now, but this has a negligible effect, and for
consistency we refrain from doing so.}. However, the sequence
$\Gamma^{M_u}$ is now selected uniformly over an alphabet
$\brk{(|\m{X}|-1)s}$, where $s=s(n)$ corresponds to a larger
number of update bits, and is defined in step (\ref{item:updateX})
below. Again, apart from active feedback positions the receiver
passively feeds back what it receives.

\item \textbf{Training transmission}: \label{item:trainingX} The
training estimate $\vect{p}^{\rm train}$ is given by
(\ref{eq:training_estimate}).

\item \textbf{Update transmission}: \label{item:updateX} Update
information is transmitted over the $M_u$ random positions
determined by $\Lambda^{b_p}$. The uncoded update information
includes the type (symbol occurrences) of the noise sequence over
regular positions in the previously accepted block, the index of
the message interval w.r.t. the interval partitioning of the
empirical posterior at the end of that block, and one ambiguity
resolving symbol. Using a binary representation, the total number
of uncoded update bits is no more than
\begin{equation}\label{eq:update_bits_num_nonbinary}
\left\lceil\log{b^{(|\m{X}|-1)}}\right\rceil +
\left\lceil\log{(n(|\m{X}|-1)+1)}\right\rceil + 1 \leq
2(|\m{X}|-1)\lceil\log(n+1)\rceil \dfn s(n)
\end{equation}
and again we zero pad the  uncoded update bits up to the length
$s$ above, for simplicity. For a non-binary alphabet, using a
random ``repetition code'' method similar to the one used in the
binary alphabet case may result in a decoding
ambiguity\footnote{This occurs when the empirical distribution of
the noise inside the block is invariant under some cyclic shift.
Take for example the distribution $(0.4,0.1,0.4,0.1)$ over a
quaternary alphabet, in which case one cannot separate, say, the
all '0's and the all '2's repetition words, but the empirical
capacity is nevertheless positive. Even in the simple
modulo-additive DMC setting with the above noise distribution, one
would use only two inputs to attain capacity, say the first and
the second.}, and thus a different method must be used. The
sequence $\Gamma^{M_u}$ takes values over an alphabet
$\brk{(|\m{X}|-1)s}$, so we can write for any $k$, $\Gamma_k=i+js$
for some $i\in \brk{s}$ and $j\in \brk{\,|\m{X}|-1}$. Following
this representation, in the $k$th update position we send one of
the symbols from the pair $\{0,j+1\}$, where which one is
determined by the $i$th uncoded update bit\footnote{We use $'0'
\rightarrow\{0\}\,, '\hspace{-0.08cm}1'\rightarrow\{j+1\}$}. For a
suitable selection of parameters, this procedure guarantees (with
high probability) that any of the uncoded update bits is sent
several times using pairs of channel inputs with any possible
modulo-additive distance. This in turn guarantees that any bit is
resolvable with high probability via at least one of the pairs,
unless the empirical capacity of that block is close to zero.

\item \textbf{Horstein iterations with KT($b$) estimates}: Similar
to the binary alphabet case. Horstein iterations are performed
over the (random) $M_r$ regular positions determined by
$\Lambda^{b_p}$, using the most updated KT estimate of the
empirical noise distribution (over accepted blocks) available to
the receiver.

\item \textbf{Block discarding}: Same criterion as in the the
binary alphabet case, where now $\vect{p}_u$ in
(\ref{eq:discarding_criteria}) is taken to be the uniform
distribution over $\m{X}$.

\item \textbf{Update information decoding}:  For an accepted
block, the update information is decoded using $\vect{p}^{\rm
train}$ as follows. Let $B^{b_p}_{(i,j)}$ be the repetition
pattern sequence of the $i$th update bit using the inputs
$\{0,j+1\}$, as determined by $\Lambda^b,\Gamma^{M_u}$. For any
$i\in\brk{s}$ and $j\in\brk{\,|\m{X}|-1}$, the receiver calculates
the following update estimate:
\begin{equation*}
\vect{p}^{{\rm upd}}_{(i,j)}\;\dfn\;
\alpha\left(B^{b_p}_{(i,j)}\right)\vect{p}_{\rm
emp}\left(\tilde{Y}^{b_p}\samp B^{b_p}_{(i,j)}\right)=
\left(\frac{n_{i+js}\left(\Gamma^{M_u}\right)}{m\slash((|\m{X}|-1)s)}\right)\cdot\vect{p}_{\rm
emp}\left(\tilde{Y}^{b_p}\samp B^{b_p}_{(i,j)}\right)
\end{equation*}
The decoding rule for the $i$th update bit is given by
\begin{equation}\label{eq:update_dec_rule}
\max_{j\in\brk{|\m{X}|-1}}\;\Linf{\vect{p}^{\rm
train}-\vect{p}^{{\rm upd}}_{(i,j)}} \;\decision{0}{1} \;\;\tau_u
\end{equation}
where in cases of an equality a `0' is decoded. The decoded
information is used to update the KT estimate, and to store the
new identity of the message interval.

\end{enumerate}

The decoding rule used is the same one as in the binary setting,
see subsection \ref{subsec:full_scheme}. In the following section,
we analyze the performance of the described scheme, proving it is
universal for the family $\mathscr{M}_{\sstm{X}}$.

\section{Analysis}\label{sec:analysis}
In this section we analyze the performance of the finite-alphabet
finite-horizon scheme presented in subsection
\ref{subsec:modulo-additive}, and show it achieves the empirical
capacity in the limit of infinite horizon, for a suitable
selection of the parameters $m(n),b(n),\tau_{\sst
d}(n),\tau_u(n)$. For brevity, the dependence of the parameters on
$n$ will usually be omitted. In the sequel, we also show how
faster convergence to the empirical capacity is obtained when
operating over noise sequence channels, and discuss the amount of
randomness generated by the scheme. The following observation
plays a key role in our subsequent derivations:
\begin{lemma}\label{lem:sampling}
For any specific block, let $\widetilde{Z}^{b_p}$, $B^{b_p}$ and
$B^{b_p}_{(i,j)}$ be the corresponding noise sequence over passive
positions, training pattern sequence, and repetition pattern
sequence for the $i$th update bit with the input pair $\{0,j+1\}$,
respectively. Then $B^{b_p}$ and $B^{b_p}_{(i,j)}$ each constitute
a causal sampling sequence for $\widetilde{Z}^{b_p}$.
\end{lemma}
\begin{proof}
See Appendix \ref{app:lemmas}. In short, the training/update
positions are i.i.d. by construction, and causal independence is
established by combining that with the Markov relations
(\ref{eq:markov_relation}) and (\ref{eq:Markov_relation_mod_add}).
\end{proof}

\subsection{Error Probability}\label{subsec:err_prob}

The only source for error in our scheme lies in the incorrect
decoding of update information in the last accepted block before
transmission is terminated, which causes the wrong message
interval to be decoded. However, for simplicity of the exposition,
we leniently assume that an error is declared whenever the update
information in any of the blocks is erroneously decoded.

The error probability is hence upper bounded by the probability of
erroneous update decoding in any of the blocks. Therefore, we now
focus on a specific block and find the corresponding update
decoding error probability, where it is emphasized that while
discarding a block has an impact on the rate, it does not
constitute an error event. The noise sequence over passive
positions in the block is denoted as before by
$\widetilde{Z}^{b_p}$. For any $i\in\brk{s}$ and
$j\in\brk{\,|\m{X}|-1}$, define
\begin{equation}\label{eq:update_est_noise}
\vect{p}_{(i,j)}\;\dfn\;
\alpha\left(B^{b_p}_{(i,j)}\right)\vect{p}_{\rm
emp}\left(\tilde{Z}^{b_p}\samp B^{b_p}_{(i,j)}\right)
\end{equation}
which is the counterpart of $\vect{p}^{\rm upd}_{(i,j)}$, yet
sampling the noise sequence rather than the output sequence over
the update positions corresponding to the $i$th update bit and the
input pair $\{0,j+1\}$. Define the following two events:
\begin{align*}
E_1\;\defn\;\left\{\Linf{\vect{p}^{\rm train}-\vect{p}_{\rm
emp}(\widetilde{Z}^{b_p})}>\tau\right\}\,,\quad
E_2\;\dfn\;\left\{\max_{i\in\brk{s}\,,j\in\brk{|\m{X}|-1}}\Linf{\vect{p}_{(i,j)}-\vect{p}_{\rm
emp}(\widetilde{Z}^{b_p})}>\tau\right\}
\end{align*}
For some $\tau(n)=\lito(1)$. We assert that for a suitable
selection of the thresholds $(\tau_{\sst d},\tau_u,\tau)$, a
necessary condition for an update decoding error in the block is
given by the event $E_1\cup E_2$. To see why this holds, let us
assume the complementary event $E_1^c\cap E_2^c$ and show it
implies no update decoding errors for a suitable thresholds
selection. If the block was discarded then surely there is no
error, so assume further the block was not discarded. Now consider
the $i$th update bit. If this bit is a `0' then the channel input
at the corresponding update positions (determined by
$B^{b_p}_{(i,j)}$) is $0\in\m{X}$, and therefore
$\vect{p}_{(i,j)}=\vect{p}^{\rm upd}_{(i,j)}$ for any
$j\in\brk{|\m{X}|-1}$. Thus in this case we have
\begin{equation*}
\Linf{\vect{p}^{{\rm upd}}_{(i,j)}-\vect{p}^{\rm train}} =
\Linf{\vect{p}_{(i,j)}-\vect{p}^{\rm train}} \leq
\Linf{\vect{p}_{(i,j)}-\vect{p}_{\rm
emp}(\widetilde{Z}^{b_p})}+\Linf{\vect{p}^{\rm
train}-\vect{p}_{\rm emp}(\widetilde{Z}^{b_p})} \leq 2\tau
\end{equation*}
which holds for any $j\in\brk{|\m{X}|-1}$. Therefore, if we set
$2\tau\leq\tau_u$ then the above together with the update decoding
rule (\ref{eq:update_dec_rule}) imply that the $i$th update bit is
correctly decoded. Now suppose the $i$th update bit is a `1', in
which case the channel input at the corresponding update positions
is $(j+1)\in\m{X}$, and thus $\vect{p}^{\rm upd}_{(i,j)}$ is a
\textit{cyclic right-shift} of $\vect{p}_{(i,j)}$ by $j+1$
positions. Writing $\vect{p}^{(-j)}_{\rm
emp}(\widetilde{Z}^{b_p})$ for a \textit{cyclic left-shift} of
$\vect{p}_{\rm emp}(\widetilde{Z}^{b_p})$ by $j+1$ positions, we
have the following chain of inequalities:
\begin{align}\label{eq:update_decoding_robust1}
\nonumber&\Linf{\vect{p}^{{\rm upd}}_{(i,j)}-\vect{p}^{\rm train}}
\;\stackrel{(\rm a)}{\geq}\; \Linf{\vect{p}^{{\rm
upd}}_{(i,j)}-\vect{p}_{\rm
emp}(\widetilde{Z}^{b_p})}-\Linf{\vect{p}^{\rm
train}-\vect{p}_{\rm emp}(\widetilde{Z}^{b_p})} \;\stackrel{(\rm
b)}{\geq}\; \;\Linf{\vect{p}^{{\rm upd}}_{(i,j)}-\vect{p}_{\rm
emp}(\widetilde{Z}^{b_p})}-\tau
\\
& \;= \;\left(\Linf{\vect{p}^{{\rm upd}}_{(i,j)}-\vect{p}_{\rm
emp}(\widetilde{Z}^{b_p})}+\Linf{\vect{p}_{(i,j)}-\vect{p}_{\rm
emp}(\widetilde{Z}^{b_p})}\right)-\Linf{\vect{p}_{(i,j)}-\vect{p}_{\rm
emp}(\widetilde{Z}^{b_p})}-\tau
\\
\nonumber&\;\stackrel{(\rm
c)}{\geq}\;\left(\Linf{\vect{p}_{(i,j)}-\vect{p}^{(-j)}_{\rm
emp}(\widetilde{Z}^{b_p})}+\Linf{\vect{p}_{(i,j)}-\vect{p}_{\rm
emp}(\widetilde{Z}^{b_p})}\right)-2\tau \;\stackrel{(\rm
d)}{\geq}\; \;\Linf{\vect{p}^{(-j)}_{\rm
emp}(\widetilde{Z}^{b_p})-\vect{p}_{\rm
emp}(\widetilde{Z}^{b_p})}-2\tau
\end{align}
In transition (a) we used the triangle inequality for the
$\m{L}_\infty$ norm, transition (b) holds since we assume $E_1^c$,
in (c) we use $E_2^c$, and also replace a cyclic right-shift of
one vector with a corresponding cyclic left-shift of the other
vector inside the first $\m{L}_\infty$ norm term. Finally, the
triangle inequality is used once again in (d).

We can now maximize both sides of
(\ref{eq:update_decoding_robust1}) over $j\in\brk{|\m{X}|-1}$, to
obtain
\begin{align}\label{eq:update_decoding_robust2}
\max_{j\in\brk{|\m{X}|-1}}\Linf{\vect{p}^{{\rm
upd}}_{(i,j)}-\vect{p}^{\rm train}} &\;\geq\;
\max_{j\in\brk{|\m{X}|-1}}\Linf{\vect{p}^{(-j)}_{\rm
emp}(\widetilde{Z}^{b_p})-\vect{p}_{\rm
emp}(\widetilde{Z}^{b_p})}-2\tau
\\
\nonumber&\;\stackrel{(\rm a)}{=}\; \max\left(\vect{p}_{\rm
emp}(\widetilde{Z}^{b_p})\right)-\min\left(\vect{p}_{\rm
emp}(\widetilde{Z}^{b_p})\right)-2\tau \;\stackrel{(\rm b)}{>}\;
\Linf{\vect{p}_{\rm emp}(\widetilde{Z}^{b_p})-\vect{p}_u}-2\tau
\end{align}
Where $\max(\cdot),\min(\cdot)$ return the maximal and minimal
element of the vector argument, respectively. Transition (a) holds
since the maximization is over the $\m{L}_\infty$ distance between
a vector and all its cyclic shifts, and for (b) to hold with a
strict inequality we further assume that $\vect{p}_{\rm
emp}(\widetilde{Z}^{b_p})$ is not precisely uniform, which is
satisfied by setting $\tau<\tau_{\sst d}$, since
\begin{align*}
\tau_{\sst d} \leq \Linf{\vect{p}^{\rm train}-\vect{p}_u} \leq
\Linf{\vect{p}_{\rm emp}(\widetilde{Z}^{b_p})-\vect{p}_u}
+\Linf{\vect{p}^{\rm train}-\vect{p}_{\rm
emp}(\widetilde{Z}^{b_p})} \leq \Linf{\vect{p}_{\rm
emp}(\widetilde{Z}^{b_p})-\vect{p}_u}+\tau
\end{align*}
Finally, combining the above with
(\ref{eq:update_decoding_robust2}) we obtain
\begin{align*}
\max_{j\in\brk{|\m{X}|-1}}\Linf{\vect{p}^{{\rm
upd}}_{(i,j)}-\vect{p}^{\rm train}} \;>\; \tau_{\sst d} -3\tau
\end{align*}
If we set $\tau_{\sst d}-3\tau\geq\tau_u$ then the above together
with the update decoding rule (\ref{eq:update_dec_rule}) imply
that the $i$th update bit is correctly decoded in this case as
well. Therefore, we now set
\begin{equation}\label{eq:thresholds_setting}
\tau_u = 2\tau\,,\quad \tau_{\sst d} = 5\tau
\end{equation}
and continue our analysis henceforth depending on the parameter
$\tau=\tau(n)$. As we have just seen, this selection guarantees
that the event $E_1\cup E_2$ is indeed a necessary condition for
an update decoding error within the block.

Let us now bound the probability of the event $E_1$. To that end,
we note that Lemma \ref{lem:sampling} together with the
$\alpha$-normalization used in the definition of the training
estimate, facilitate the use of Lemma \ref{lem:azuma_sampling}.
Since the training pattern sequence has a marginal distribution
$\sim{\rm Ber}(q)$ with $q=\frac{m}{b_p}$, we obtain
\begin{equation*}
\Prob(E_1) = \Prob\left(\Linf{\vect{p}^{\rm train}-\vect{p}_{\rm
emp}(\widetilde{Z}^{b_p})}>\tau\right)  \leq
2|\m{X}|\exp\left(-\frac{b_p\tau^2\left(\frac{m}{b_p}\right)^2}{2}\right)
\leq 2|\m{X}|\exp\left(-\frac{1}{2}\,\tau^2m^2b^{-1}\right)
\;\dfn\; \eps_1^{(1)}(n)
\end{equation*}

Bounding the probability of the event $E_2$ is similar, and Lemma
\ref{lem:sampling} together with (\ref{eq:update_est_noise})
facilitate the use of Lemma \ref{lem:azuma_sampling} for any of
the repetition pattern sequences. These sequences all have a
marginal distribution $\sim{\rm Ber}(q)$ with
$q=\frac{m}{b_p(|{\sstm{X}}|-1)s}\,$ where
$s=2(|\m{X}|-1)\lceil\log(n+1)\rceil$ was given in
(\ref{eq:update_bits_num_nonbinary}). Using Lemma
\ref{lem:azuma_sampling} and applying the union bound over all
update bits and input pairs (i.e., all repetition pattern
sequences) leads to
\begin{align*}
\Prob(E_{2}) \;&\leq\;
\sum_{i\in\brk{s},j\in\brk{|\m{X}|-1}}\Prob\left(\Linf{\vect{p}_{(i,j)}-\vect{p}_{\rm
emp}(\widetilde{Z}^{b_p})}>\tau\right) \;\leq\; (|\m{X}|-1)s\cdot
2|\m{X}|\exp\left(-\frac{\tau^2m^2b^{-1}}{2(|\m{X}|-1)^2s^2}\right)
\\
\;&\leq\;
4|\m{X}|(|\m{X}|-1)^2\lceil\log(n+1)\rceil\exp\left(-\frac{\tau^2m^2b^{-1}}{8(|\m{X}|-1)^4\lceil\log(n+1)\rceil^2}\right)
\;\dfn\; \eps_1^{(2)}(n)
\end{align*}

So far, we have established that the probability of an update
decoding error in any given block is upper bounded by
$\Prob(E_1\cup E_2)\leq \eps_1^{(1)}(n)+\eps_1^{(2)}(n)$. Using
the union bound over the blocks and the fact that
$\eps_1^{(1)}(n),\eps_1^{(2)}(n)$ do not depend on the message
point or the channel, we obtain a uniform upper bound for the
error probability achieved by the scheme:
\begin{equation}\label{eq:err_prob}
\sup_{\m{W}\in\mathscr{M}_{\sstm{X}},\theta_0\in[0,1)}p_e(n,\m{W},\theta_0)
\;\leq\; nb^{-1}\left(\eps_1^{(1)}(n)+\eps_1^{(2)}(n)\right)
\;\dfn\; \eps_1(n)
\end{equation}
From the expression above it is easily verified that if
$b(n),m(n),\tau(n)$ are selected such that $\tau^2 m^2b^{-1} =
\omega\left(\log^2{n}\log{\left(b^{-1}n\log{n}\right)}\right)$,
then $\eps_1(n)\rightarrow 0$ and the error probability tends to
zero uniformly as desired. However, since this is a variable rate
scheme, a low probability of error is not enough since while not
making an error indicates we have correctly decoded bits, it does
not indicate how many.

\subsection{Rate}\label{subsec:dec_rate}

Due to the inherent randomness generated by the transmission
scheme and the possibly random actions of the channel, the rate
achieved by the scheme is random\footnote{Note that even in the
case of an individual noise sequence, the rate is still random due
to training/update randomization.}. In this section we show that
this rate is arbitrarily close to the empirical capacity of the
channel, with probability that tends to one.

At the first stage of the proof, we look only at regular positions
(which are used for Horstein iterations), and analyze the rate
w.r.t. these channel uses only. Later, we make the necessary
adjustments taking into account the negligible effect of
non-regular positions as well. For an accepted block, the number
of regular positions is in the range $(b_{\rm min},b_{\rm max})$,
where $b_{\rm min} = b_p - 4m\,,b_{\rm max} = b_p - m$. Let
$n^{\rm reg}$ be the (random) total number of regular positions
over the entire transmission period, and let
$\beta\in[\hspace{0.01cm}0,1]$ denote the (random) fraction of
these positions that are accepted (namely, reside in accepted
blocks). Communications (via Horstein iterations) take place only
on accepted regular positions, namely over $\beta n^{\rm reg}$
channel uses. The KT estimates used by the decoder are updated at
varying intervals, but these intervals do not exceed $2b_{\rm
max}$ (measured relative to the sequence of accepted regular
positions). Hence, the estimator used in effect is a KT($2b_{\rm
max}$) estimator over a sequence of length $\beta n^{\rm reg}$.

Define $V_0$ to be the event where no update decoding errors have
occurred, and let $V_1$ be the event where none of the blocks were
discarded due to a too small or too large selection of $M_t,M_u$
made by the receiver. Given $V_0$, the KT estimates are based on
noiseless observations of the noise sequence. Given $V_1$, we have
$nb^{-1}b_{\rm min}\leq n^{\rm reg} \leq nb^{-1}b_{\rm max}$. Let
$R^{\rm reg}$ be the (random) decoding rate measured over regular
channel positions only (including both accepted and discarded
blocks), and denote by $\vect{p}^{\rm reg}_{a}$ the (random)
empirical distribution of the noise sequence over accepted regular
positions (entire transmission period). Using
(\ref{eq:rate_pure_fin_alph}) and substituting $n\rightarrow\beta
n^{\rm reg}$ and $b\rightarrow 2b_{\rm max}$ we have that given
$V_0\cap V_1$
\begin{align}\label{eq:rate_beta}
\nonumber R^{\rm reg} &\geq
\beta\left(\log{|\m{X}|}-H(\vect{p}^{\rm
reg}_{a})-K_2\frac{2|\m{X}|b_{\rm max}\log{(\beta n^{\rm
reg}})}{\beta n^{\rm reg}}\right) =
\beta\big{(}\log{|\m{X}|}-H(\vect{p}^{\rm reg}_{a})\big{)} -
K_2\frac{2|\m{X}|b_{\rm max}\log{(\beta n^{\rm reg}})}{n^{\rm
reg}}
\\
& \geq\; R_{\sst\beta} -2|\m{X}|K_2\cdot\frac{b}{b_{\rm
min}}\cdot\frac{b\log{n}}{n}
\end{align}
where $R_{\sst\beta} \dfn
\beta\big{(}\log{|\m{X}|}-H(\vect{p}^{\rm reg}_{a})\big{)}$.

We now focus on the principal rate term $R_{\sst\beta}$. As
already mentioned, due to the concavity of the entropy it is
expected that discarding blocks will only increase the achieved
rate with high probability, as discarded blocks usually have a low
empirical capacity. Therefore, we would like to seek conditions
under which $R_{\sst\beta}$ is minimized by $\beta=1$ (no
discarded blocks), and later show that these conditions are
satisfied with high probability. Denote by $\vect{p}^{\rm reg}$
and $\vect{p}^{\rm reg}_{d}$ the (random) empirical distributions
of noise sequence over all regular positions, and over regular
positions inside discarded blocks only, respectively. These
distributions together with $\vect{p}^{\rm reg}_{a}$ satisfy
\begin{equation*}
\vect{p}^{\rm reg} = \beta \vect{p}^{\rm reg}_{a} +
(1-\beta)\vect{p}^{\rm reg}_{d}
\end{equation*}
Extracting $\vect{p}^{\rm reg}_{a}$ and substituting into the
expression for $R_{\sst\beta}$ yields
\begin{equation*}
R_{\sst\beta} = \beta\Big{(}\log{|\m{X}|}-H(\vect{p}^{\rm reg}_{d}
+\beta^{-1}(\vect{p}^{\rm reg}-\vect{p}^{\rm reg}_{d})\Big{)}
\end{equation*}
Note that for any given values of $\vect{p}^{\rm reg}$ and
$\vect{p}^{\rm reg}_{d}$, $R_{\sst\beta}$ is defined only for
values of $\beta$ large enough such that $\vect{p}^{\rm reg}_{d}
+\beta^{-1}(\vect{p}^{\rm reg}-\vect{p}^{\rm reg}_{d})$ is still a
probability vector. Now, if the derivative of $R_{\sst\beta}$
w.r.t. $\beta$ were to be non-positive over all the range of
permissible $\beta$, then $R_{\sst\beta}$ would be minimized by
$\beta=1$. We would therefore like to derive a condition for the
non-positivity of the derivative.
\begin{lemma}\label{lem:derivative}
For any given $\vect{p}^{\rm reg}, \vect{p}^{\rm reg}_{d}$ and
corresponding permissible $\beta$,
\begin{equation*}
\frac{\partial R_{\sst\beta}}{\partial\beta} \;\leq\;
\log{|\m{X}|}-H(\vect{p}_d^{\rm reg})-D\left (\vect{p}_d^{\rm
reg}\,\|\,\vect{p}^{\rm reg}\,\right)
\end{equation*}
\end{lemma}
\begin{proof}
See Appendix \ref{app:lemmas}.
\end{proof}
Based on the Lemma above, the following chain of inequalities
provide a $\m{L}_\infty$-type upper bound on the derivative:
\begin{align*}
\frac{\partial R_{\sst\beta}}{\partial\beta} &\;\leq\;
\log{|\m{X}|}-H(\vect{p}_d^{\rm reg})-D\left (\vect{p}_d^{\rm
reg}\,\|\,\vect{p}^{\rm reg}\,\right) \;\stackrel{(\rm a)}{\leq}\;
|\m{X}|\log{|\m{X}|}\Linf{\vect{p}_d^{\rm reg}-\vect{p}_u} -
\frac{1}{2\ln{2}}\left\|\vect{p}_d^{\rm reg}-\vect{p}^{\rm
reg}\right\|_1^2
\\
&\;\stackrel{(\rm b)}{\leq}\;
|\m{X}|\log{|\m{X}|}\Linf{\vect{p}_d^{\rm reg}-\vect{p}_u} -
\frac{1}{2\ln{2}}\Linf{\vect{p}_d^{\rm reg}-\vect{p}^{\rm reg}}^2
\\
&\;\stackrel{(\rm c)}{\leq}\;
|\m{X}|\log{|\m{X}|}\Linf{\vect{p}_d^{\rm reg}-\vect{p}_u} -
\frac{1}{2\ln{2}}\left(\Linf{\vect{p}_d^{\rm
reg}-\vect{p}_u}-\Linf{\vect{p}^{\rm reg}-\vect{p}_u}\right)^2
\end{align*}
where $\|\cdot\|_1$ is the $\m{L}_1$ norm. Transition (a) is due
to Pinsker's inequality for the relative entropy \cite{cover} and
the $\m{L}_\infty$ bound for the entropy (Lemma
\ref{lem:entopry_Linf_bound}), transition (b) holds since the
$\m{L}_1$ norm dominates the $\m{L}_\infty$ norm, and in (c) we
used the triangle inequality. Thus, a sufficient condition for
$\frac{\partial R_{\sst\beta}}{\partial\beta}\leq 0$ is given by
\begin{equation}\label{eq:preg_nonpos_deriv_cond0}
\Big{|}\Linf{\vect{p}^{\rm reg}-\vect{p}_u}-\Linf{\vect{p}_d^{\rm
reg}-\vect{p}_u}\Big{|} \;\geq\;
\sqrt{2\ln{2}\cdot|\m{X}|\log{|\m{X}|}}\cdot\Linf{\vect{p}_d^{\rm
reg}-\vect{p}_u}^{\frac{1}{2}}
\end{equation}
Practicing some algebra, it is easily verified\footnote{Using
$\Linf{\vect{p}_d^{\rm reg}-\vect{p}_u}\leq 1$, $\ln{2}<1$,
$|\m{X}|\geq 2$ and $x\log{x}<\frac{x^2}{2}$ for $x=2,3,\ldots$}
that a sufficient condition for (\ref{eq:preg_nonpos_deriv_cond0})
to hold is given by
\begin{equation}\label{eq:preg_nonpos_deriv_cond}
\Linf{\vect{p}^{\rm reg}-\vect{p}_u} \;\geq\; 2|\m{X}|\cdot
\Linf{\vect{p}_d^{\rm reg}-\vect{p}_u}^{\frac{1}{2}}
\end{equation}
and so given (\ref{eq:preg_nonpos_deriv_cond}) we have
$R_{\sst\beta}\geq R_{\sst\beta=1} = \log{|\m{X}|}-H(\vect{p}^{\rm
reg})$. Using (\ref{eq:rate_beta}), it is therefore established
that (\ref{eq:preg_nonpos_deriv_cond}) together with $V_0\cap V_1$
imply that
\begin{equation}\label{eq:Rreg_bound}
R^{\rm reg} \;\geq\; \log{|\m{X}|}-H(\vect{p}^{\rm reg})
-2|\m{X}|K_2\cdot\frac{b}{b_{\rm min}}\cdot\frac{b\log{n}}{n}
\end{equation}

We would like to obtain a similar result involving $\vect{p}_{\rm
emp}(Z^n)$ and the rate $R_n$. To that end, let $\eta\in[0,1]$ be
the fraction of regular positions (out of $n$), and so $R_n=\eta
R^{\rm reg}$. Let $\vect{p}^{\rm nreg}$ denote the distribution of
the noise sequence over non-regular positions. Given the event
$V_1$ we have $\eta\geq\frac{b_{\rm min}}{b}$, and so
\begin{align}\label{eq:preg2pemp}
&\Linf{\vect{p}_{\rm emp}(Z^n)-\vect{p}_u} =
\Linf{\eta(\vect{p}^{\rm reg}-\vect{p}_u)+(1-\eta)(\vect{p}^{\rm
nreg}-\vect{p}_u)} \leq \eta\Linf{\vect{p}^{\rm
reg}-\vect{p}_u}+(1-\eta)
\\
\nonumber&\leq\;\Linf{\vect{p}^{\rm
reg}-\vect{p}_u}+\frac{b-b_{\rm min}}{b}
\;\leq\;\Linf{\vect{p}^{\rm
reg}-\vect{p}_u}+4mb^{-1}\left(1+2\log{b}\right) \;\leq\;
\Linf{\vect{p}^{\rm reg}-\vect{p}_u}+K_3mb^{-1}\log{b}
\end{align}
for some $K_3>0$, where we have used the convexity of the norm,
and the fact that $\Linf{\vect{p}^{\rm nreg}-\vect{p}_u}\leq 1$.
Furthermore, using the concavity and nonnegativity of the entropy
we have that given $V_1$
\begin{equation}\label{eq:pemp2preg_ent}
H_{\rm emp}(Z^n) = H(\vect{p}_{\rm emp}(Z^n)) =
H(\eta\vect{p}^{\rm reg}+(1-\eta)\vect{p}^{\rm nreg}) \geq \eta
H(\vect{p}^{\rm reg}) \geq \frac{b_{\rm min}}{b} H(\vect{p}^{\rm
reg})
\end{equation}

Now, introduce an auxiliary parameter $\gamma(n)=\lito(1)$ chosen
to satisfy $\frac{\gamma}{2}>K_3mb^{-1}\log{b}$, which is made
feasible by requiring $mb^{-1}\log{b}=\lito(1)$. Define the events
\begin{equation}\label{eq:events_V23}
V_2 \;\dfn\; \left\{2|\m{X}|\Linf{\vect{p}_d^{\rm
reg}-\vect{p}_u}^{\frac{1}{2}}\leq
\frac{\gamma}{2}\right\}\,,\qquad
V_3\;\dfn\;\left\{\Linf{\vect{p}_{\rm emp}(Z^n)-\vect{p}_u} \geq
\gamma\right\}
\end{equation}

Now, using (\ref{eq:preg2pemp}) it is readily verified that
$V_1\cap V_2\cap V_3$ implies (\ref{eq:preg_nonpos_deriv_cond}),
and therefore $\bigcap_{i=0}^{3}V_i$ implies
(\ref{eq:Rreg_bound}). Furthermore, $V_1$ implies
(\ref{eq:pemp2preg_ent}). Putting (\ref{eq:Rreg_bound}) and
(\ref{eq:pemp2preg_ent}) together, we establish that given
$\bigcap_{i=0}^{3}V_i$,
\begin{align}\label{eq:Rn_bound}
\nonumber R_n &= \eta R^{\rm reg} \geq \frac{b_{\rm min}}{b} \cdot
R^{\rm reg} \geq \frac{b_{\rm min}}{b}
\left(\log{|\m{X}|}-\frac{b}{b_{\rm min}} H_{\rm emp}(Z^n)
-2|\m{X}|K_2\cdot\frac{b}{b_{\rm
min}}\cdot\frac{b\log{n}}{n}\right)
\\
&\;\geq\; \log{|\m{X}|}-H_{\rm emp}(Z^n) - \log{|\m{X}|}\cdot
K_4\frac{m\log{b}}{b} - 2|\m{X}|K_2\frac{b\log{n}}{n}
\end{align}
for some $K_4>0$. Moreover, given $V_3^c$ the $\m{L}_\infty$ bound
for the entropy (Lemma \ref{lem:entopry_Linf_bound}) yields
\begin{equation}\label{eq:emp_cap_give_V3c}
\log|\m{X}|-H_{\rm emp}(Z^n) \leq \gamma\,|\m{X}|\log|\m{X}|
\end{equation}
which enables us to remove the dependence on the event $V_3$ by
incorporating the above into the redundancy term. Namely, since
$\{\bigcap_{i=0}^{3}V_i\cup V_3^c\}\supseteq
\bigcap_{0=1}^{2}V_i$, we can combine (\ref{eq:Rn_bound}) and
(\ref{eq:emp_cap_give_V3c}) and the definition of the empirical
capacity, to obtain
\begin{align}
\Prob\Big{(}R_n(\m{W},\theta_0) \;\geq\; C^{\,\rm
emp}_n(\m{W},\theta_0) - \eps_2(n)\Big{)}  \geq
\Prob\left(\bigcap_{i=0}^{2}V_i\right)
\end{align}
where
\begin{equation}\label{eq:true_rate_bound}
\eps_2(n) \;\defn\; \log{|\m{X}|}\cdot K_4\frac{m\log{b}}{b} +
2|\m{X}|K_2\frac{b\log{n}}{n} + \gamma\,|\m{X}|\log|\m{X}|
\end{equation}

To conclude the rate analysis, we need to lower bound the
probability of $\bigcap_{0=1}^{2}V_i$ and set $\gamma$ as a
function of the scheme parameters. While analyzing the error
probability, we have already established that $\Prob(V_0^c)\leq
\eps_1(n)$. We note that $\Prob(V_1^c)$ is simply upper bounded by
the event where at least one of $2nb^{-1}$ Binomial r.v.s $\sim
B(b_p,\frac{m}{b_p})$ (the $M_t,M_u$ of all the blocks) deviates
by more than $\frac{m}{2}$ from its expected value. Applying (say)
the Hoeffding inequality \cite{hoeffding} and then the union
bound, we obtain
\begin{equation}\label{eq:prob_V1c}
\Prob(V_1^c) \leq 4nb^{-1}\exp\left(-\frac{2(m\slash
2)^2}{b_p}\right) \leq
4nb^{-1}\exp\left(-\frac{1}{2}\,m^2b^{-1}\right) \;\dfn\;
\eps_3^{(1)}(n)
\end{equation}

For $V_2$, we have $\widetilde{V}_2\subseteq V_2$ where
$\widetilde{V}_2^c$ is the event where at least one discarded
block has an empirical distribution $\vect{q}^{\rm reg}$ over
regular positions which does not satisfy the condition defining
the event $V_2$, namely
\begin{equation*}
2|\m{X}|\Linf{\vect{q}^{\rm reg}-\vect{p}_u}^{\frac{1}{2}} >
\frac{\gamma}{2}
\end{equation*}
We would like to obtain a corresponding necessary condition on the
deviation from uniformity of the training estimate, the
probability of which we can then bound. Using norm properties, it
is easily verified that
\begin{equation*}
\Big{|}\|\vect{p}_{\rm
emp}(\widetilde{Z}^{b_p})-\vect{p}_u\|_\infty- \Linf{\vect{q}^{\rm
reg}-\vect{p}_u}\Big{|} \leq \frac{2\cdot 4m}{b_{\rm min}} =
\frac{8m}{b-8m\lceil\log{b}\rceil-4m} \leq K_5mb^{-1}
\end{equation*}
for some $K_5>0$, where $\vect{p}_{\rm emp}(\widetilde{Z}^{b_p})$
is the corresponding empirical distribution over all passive
positions in the block. Hence, we have
$\widehat{V}_2\subseteq\widetilde{V}_2\subseteq V_2$ where
$\widehat{V}_2^c$ is the event where for at least one block with
an empirical distribution over passive positions $\vect{p}_{\rm
emp}(\widetilde{Z}^{b_p})$ and training estimate $\vect{p}^{\rm
train}$, simultaneously satisfies
\begin{align*}
\Linf{\vect{p}_{\rm emp}(\widetilde{Z}^{b_p})-\vect{p}_u} >
\left(\frac{\gamma}{4|\m{X}|}\right)^2-K_5mb^{-1}\,,\quad
\Linf{\vect{p}^{\rm train}-\vect{p}_u} < \tau_{\sst d} = 5\tau
\end{align*}
Hence, using the triangle inequality a necessary condition for
$\widehat{V}_2^c$ is for the training estimate in some block to
deviate by at least
\begin{equation}\label{eq:train_dev}
\Linf{\vect{p}^{\rm train}-\vect{p}_{\rm
emp}(\widetilde{Z}^{b_p})}
> \left(\frac{\gamma}{4|\m{X}|}\right)^2-K_5mb^{-1}-5\tau
\end{equation}
We would now like to set $\gamma$ so that the right-hand-side of
the above is strictly positive, and such that
$\frac{\gamma}{2}>K_3mb^{-1}\log{b}$ which was previously
required, is also satisfied. This is obtained by setting $\gamma$
to
\begin{equation}\label{eq:gamma_set}
\left(\frac{\gamma}{4|\m{X}|}\right)^2-K_5mb^{-1}-5\tau =
K_6\left(\tau+mb^{-1}\log{b}\right)
\end{equation}
with $K_6>0$ large enough. Using Lemma \ref{lem:azuma_sampling}
and the union bound over blocks, we get
\begin{align*}
\Prob(V_2^c) &\;\leq\; \Prob(\widehat{V}_2^c) \;\leq\;
nb^{-1}\Prob\left(\Linf{\vect{p}^{\rm train}-\vect{p}_{\rm
emp}(\widetilde{Z}^{b_p})} >
K_6\left(\tau+mb^{-1}\log{b}\right)\right)
\\
&\;\leq\;
2|\m{X}|nb^{-1}\exp\left(-\frac{1}{2}K_6^2\left(mb^{-1}\log{b}+\tau\right)^2m^2b^{-1}\right)
\;\dfn\; \eps_3^{(2)}(n)
\end{align*}
and finally,
\begin{equation}\label{eq:_redundancy_term}
\Prob\left(\bigcap_{i=0}^{2}V_i\right) \geq
1-\sum_{i=0}^2\Prob(V_i^c) \geq 1-\left(\eps_1(n) +
\eps_3^{(1)}(n)+\eps_3^{(2)}(n)\right) \;\dfn\; \eps_3(n)
\end{equation}
Note that $\eps_3(n) = \bigo(\eps_1(n))$ and hence
$\eps_3(n)\rightarrow 0$ under the same condition provided for the
error probability in subsection \ref{subsec:err_prob}.

Summarizing, we have found that a rate $\eps_2(n)$ close to the
empirical capacity (eq. (\ref{eq:true_rate_bound}) with $\gamma$
given in (\ref{eq:gamma_set})) is achieved with probability at
least $1-\eps_3(n)$ (eq. (\ref{eq:_redundancy_term})), and an
error probability no larger than $\eps_1(n)$ (eq.
(\ref{eq:err_prob})). In passing, we have also described a set of
constraint on the asymptotic behavior of the scheme parameters
that are sufficient to guarantee that
$\eps_1(n),\eps_2(n),\eps_3(n)\rightarrow 0$. In the next
subsection we summarize these constraints, and show that there
exist (many) selections of scheme parameters for which they are
satisfied.

\subsection{Parameter Selection and Asymptotic Behavior}\label{subsec:param_sel}
There are many different selections of the scheme
parameters\footnote{Recall that the thresholds $\tau_{\sst
d}(n),\tau_u(n)$ are determined by $\tau(n)$, as given in
(\ref{eq:thresholds_setting}).} $b(n),m(n),\tau(n)$ which allow
all the convergence parameters $\eps_1(n),\eps_2(n),\eps_3(n)$ to
become asymptotically negligible, and result in various trade-offs
between them. The following is the set of sufficient asymptotic
conditions to that end, derived directly from the discussion in
the two previous subsections:
\begin{align}\label{eq:param_cond}
\begin{array}{lllll}(\rm I)& \tau = \lito(1) && (\rm IV)& bn^{-1}\log{n}= \lito(1)
\\
(\rm II)& mb^{-1}\log{b}= \lito(1) && (\rm V)& \tau^2 m^2b^{-1} =
\omega\left(\log^2{n}\log{\left(b^{-1}n\log{n}\right)}\right)
\end{array}
\end{align}
Note that the above conditions also imply that
$b=\omega(\log{n})$, which was an assumption made when computing
the number of update bits. Under (\ref{eq:param_cond}), the
following asymptotical behavior for the convergence parameters is
achievable:
\begin{center}
\begin{tabular}[c]{|l|c|}
\hline & \\[-7pt] Error probability $\eps_1(n)$& $-\log\eps_1(n)     =
\Omega\left(\frac{\tau^2m^2b^{-1}}{\log^2n}\right)$
\\[5pt]
\hline & \\[-7pt] Target redundancy $\eps_2(n)$& $\eps_2(n) =
\bigo(bn^{-1}\log{n})+\bigo\left(\sqrt{\tau+mb^{-1}\log{b}}\,\right)$
\\[5pt]
\hline & \\[-7pt] Redundancy exceeding probability $\eps_3(n)$& $-\log\eps_3(n) =
\Omega\left(\frac{\tau^2m^2b^{-1}}{\log^2n}\right)$
\\[5pt]
\hline
\end{tabular}
\end{center}
To demonstrate that the sufficient conditions
(\ref{eq:param_cond}) can be met, let us specifically set the
parameters to
\begin{equation}\label{eq:poly_params}
b(n) = n^{a_0}\,, m(n) = n^{a_1}\,, \tau(n) = n^{-a_2}
\end{equation}
for some positive constants $a_0,a_1,a_2$. The conditions then
translate into
\begin{align*}
& a_1<a_0 <1 \,,\quad a_0<2(a_1-a_2)
\end{align*}
It is easy to find many parameter selections satisfying the above
conditions, and one possible selection is given by $(a_0,a_1,a_2)
= \left(\frac{3}{4},\frac{1}{2},\frac{1}{16}\right)$.

\subsection{Noise Sequence Channels}\label{subsec:noise_seq_chan}
As already mentioned, the family of noise sequence channels
$\mathscr{N}_{\sstm{X}}$ is a subfamily of the family of
modulo-additive channels $\mathscr{M}_{\sstm{X}}$, and therefore
the analysis presented thus far specifically holds when
communications take place over an unknown member of
$\mathscr{N}_{\sstm{X}}$. Specifically, this is also true for the
special case of an individual noise sequence, which was given as a
motivating example in section \ref{sec:introduction}. However, it
turns out that when transmission takes place over the family
$\mathscr{N}_{\sstm{X}}$ it is possible to obtain better
convergence tradeoffs than when operating over
$\mathscr{M}_{\sstm{X}}$. This is achieved by the following simple
modification within each block: The sequence $\Lambda^{b_p}$ is
drawn uniformly over the type of all sequences with
\textit{exactly} $m$ training positions and \textit{exactly} $m$
update positions. The sequence $\Gamma^m$ (note that now $M_u=m$)
is drawn uniformly over the type of all sequences with a uniform
composition\footnote{We assume $m$ divides the size of the
alphabet, however using a close to uniform composition works as
well.} over the alphabet $\brk{(|\m{X}|-1)s}$.

These changes amounts to using a fixed number of training/update
positions, and using a fixed repetition code for each update bit
with each input pair, which means that position types are not
selected in an i.i.d. fashion anymore. Thus, a fully informed
adversary can now predict the type of the next position with some
accuracy, and possibly exhibit `atypical behavior'' accordingly
(say over training positions), rendering the scheme useless. For
noise sequence channels however, the noise sequence is ``generated
separately'' from the input/output sequence (in the sense
described by the Markov relation
(\ref{eq:Markov_relation_noise_seq})), hence the adversary cannot
change its strategy based in its ability to predict, which is why
the scheme can still work. The most basic example for that is the
case where the noise is an individual sequence, which is fixed at
the beginning of time and cannot adapt according to the observed
inputs/outputs.

The only derivation in the achievability proof that needs to be
modified is that of the deviation probability of a sample's
empirical distribution from the true empirical distribution, where
this sample is now uniform over a type. To this end, we can use
Lemma \ref{lem:hoeffding_sampling} (sampling without replacement)
in lieu of Lemma \ref{lem:azuma_sampling}, as the noise sequence
is now statistically independent of the (training, update)
sampling sequences. Interestingly, the exponential decay of the
deviation probability is linear in the \textit{number of samples}
(either $m$ or
$\frac{m}{2\left(|\sstm{X}|-1\right)^2\lceil\log{n+1}\rceil}$ in
our case) and does not involve the length of the sequence sampled
from ($b_p$ in our case). Therefore, the expressions for
$\eps_i(n)$ for noise sequence channels is essentially given by
exchanging the expressions $b^{-1}m^2\rightarrow m$ and
$\lceil\log(n+1)\rceil^2\rightarrow \lceil\log(n+1)\rceil$ in all
the exponents (up to constant factors). A set of sufficient
conditions for communications over $\mathscr{N}_{\sstm{X}}$ is
given by
\begin{align}\label{eq:param_cond_noise}
\begin{array}{lllll}(\rm I)& \tau = \lito(1) && (\rm V)& \tau^2 m =
\omega\left(\log{n}\log\left(b^{-1}n\log{n}\right)\right)
\\
(\rm II)& mb^{-1}\log{b}= \lito(1) && (\rm VI)&b =
\omega\left(\log{n}\right)
\\
(\rm III)& bn^{-1}\log{n}= \lito(1) && &
\end{array}
\end{align}
and the corresponding asymptotical behavior of the convergence
parameters is given by
\begin{center}
\begin{tabular}[c]{|l|c|}
\hline & \\[-7pt] Error probability $\eps_1(n)$ & $-\log\eps_1(n) = \Omega\left(\frac{\tau^2m}{\log{n}}\right)$
\\[5pt]
\hline & \\[-7pt] Target redundancy $\eps_2(n)$ & $\eps_2(n) =
\bigo(bn^{-1}\log{n})+\bigo\left(\sqrt{\tau+mb^{-1}\log{b}}\,\right)$
\\[5pt]
\hline & \\[-7pt] Redundancy exceeding probability $\eps_3(n)$ & $-\log\eps_3(n) = \Omega\left(\frac{\tau^2m}{\log{n}}\right)$
\\[5pt]
\hline
\end{tabular}
\end{center}

Finally, note that moving from i.i.d. sampling to fixed-size
sampling without replacement is fundamental for reaping this
performance gain in noise sequence channels, since even when the
noise is independent of the sampling sequence, the tail of the
binomial distribution renders i.i.d. sampling inferior.

\subsection{Randomness Resources}\label{subsec:rand_resource}
Randomness is a key element in achieving the empirical capacity.
Let us examine just how many common random bits are consumed by
our scheme. The receiver generates $\bigo(nb^{-1}m\log{b})$ random
bits over the entire transmission period. Under the parameter
constraints (\ref{eq:param_cond}) or (\ref{eq:param_cond_noise})
this amount is sub-linear in $n$, as otherwise it could not be
accommodated by feedback. It is easily verified that these
conditions imply that for any $\eps>0$, the following amount of
randomness is sufficient for achieving the empirical capacity, for
the different channel families\footnote{The parameters $b,m$ are
provided, it is readily verified that $\tau$ can be set to satisfy
the required conditions.}:
\begin{small}
\begin{center}
\begin{tabular}[c]{|c|c|c|c|}
\hline &&& \\[-7pt] Family of Channels & Random Bits &
Generation Mechanism & Parameters
\\[5pt]
\hline &&& \\[-7pt] Noise Sequence $\mathscr{N}_{\sstm{X}}$ &
$\bigo(\log^{3+\eps}{n})$& Feedback &
$b=\Omega\left(\frac{n}{(\log{n})^{1+\frac{\eps}{2}}}\right),
m=\bigo\left({(\log{n})^{1+\frac{\eps}{2}}}\right)$
\\[5pt]
\hline &&& \\[-7pt] Modulo-Additive $\mathscr{M}_{\sstm{X}}$ &
$\bigo(\sqrt{n}\log^{2+\eps}{n})$& Feedback &
$b=\Omega\left(\frac{n}{(\log{n})^{1+\frac{\eps}{2}}}\right),
m=\bigo\left(\sqrt{n(\log{n})^{1+\eps}}\right)$
\\[5pt]
\hline &&& \\[-7pt] General Causal $\mathscr{C}_{\sstm{X}}$ &
$\bigo(n)$& Common Randomness & Any feasible selection
\\[5pt]
\hline
\end{tabular}
\end{center}
\end{small}
Interestingly, the randomness resources consumed are significantly
reduced when operating over $\mathscr{N}_{\sstm{X}}$, i.e., over
noise sequence channels. Thus, in a sense it seems that when
working against an $\mathscr{M}_{\sstm{X}}$ adversary most of the
randomness resources are dedicated to ``decoupling'' its actions
from the channel inputs/outputs, and only a negligible amount of
randomness is used for combating the ``noise effect'' itself. In
the most general case of communication over
$\mathscr{C}_{\sstm{X}}$, a much larger amount of random bits
(mainly used for dithering) cannot be accommodated by the feedback
link, and an external common randomness source is required.

\section{Summary and Discussion}\label{sec:summary}

The universal communication problem over an unknown discrete
channel with noiseless feedback was addressed. An extreme channel
uncertainty model was considered, where the channel law is unknown
to both transmitter and receiver, and may vary arbitrarily from
symbol to symbol depending on previous inputs and outputs,
possibly in an adversarial fashion. Although in such a general
setting no positive rate can be guaranteed in advance, it was
constructively shown that reliable communications at a variable
rate that corresponds to the empirical goodness of the channel,
can be attained. As a measure for this empirical goodness, the
\textit{empirical capacity} of the channel was defined as the
capacity of an equivalent memoryless modulo-additive channel, with
an additive noise marginal distribution given by the empirical
distribution of a noise sequence realized by channel actions
throughout transmission. An explicit sequential transmission
scheme was then described, and shown to achieve rates arbitrarily
close to the empirical capacity with probability approaching one,
independent of the actual channel in use and uniformly over the
message set. For the special case of individual noise sequence
channels, the scheme is universal in the sense of successfully
competing with any fixed-rate transmission scheme that knows the
empirical distribution of the noise sequence in advance.

Achieving the empirical capacity requires randomization. This is
especially evident in the case of an individual noise sequence
channel, where it is well known that deterministic coding schemes
cannot attain the empirical capacity uniformly over the message
set in general, even if the empirical distribution of the noise
sequence is given in advance. Consequently, the described scheme
requires the generation of common random bits shared by the
transmitter and the receiver. In the most general setting,
$\bigo(n)$ random bits are used by the scheme, a quantity
requiring an external source of common randomness available to the
terminals. However, if the channel law is known to be
modulo-additive at any time instant (but otherwise arbitrary
varying, depending on previous inputs/outputs), only
$\bigo(\sqrt{n}\log^{2+\eps}{n})$ random bits are sufficient for
any $\eps>0$, an amount that can be generated exclusively via
feedback at no asymptotical cost. Furthermore, in the special case
of noise sequence channels (where the channel is completely
defined by the noise sequence) the scheme exhibits improved
performance in terms of error probability and redundancy, and the
amount of common randomness is further reduced to merely
$\bigo(\log^{3+\eps}{n})$ random bits, which again can be produced
by feedback alone.

The tradeoff between error probability and transmission time
attained by the scheme is sub-exponential in $n$. This is to be
expected, since the actual channel over which communications take
place might be (say) a BSC, in which case the empirical capacity
converges to the channel capacity a.s., so one cannot hope to
universally obtain a positive error exponent when operating at the
empirical capacity. However, if one is willing to give up a
constant portion of the empirical capacity then it is plausible
that a positive error exponent could be universally attained, yet
we were unable to adapt our scheme to that end. In order to make
the errors due to training estimate deviations vanish
exponentially with $n$, a linear number of training positions must
be set, which in the finite-horizon setting implies a constant
number of blocks. This however results in an excessively slow
update rate for the KT estimate which prohibits any positive rate
from being attained, and it therefore seems that an altogether
different approach is required.

In this paper the discussion was limited to a memoryless
modulo-additive model, where universality is sought w.r.t. the
marginal empirical distribution of the realized noise sequence. In
part two of this work \cite{empirical_capacity_partII}, the
concepts presented here are further developed to encompass more
general models for channel actions. Specifically, we discuss
models that take into account \textit{empirical dependencies}
between the channel actions and the input, and exploit
\textit{empirical memory} within consecutive channel actions. This
approach will facilitate universality w.r.t. higher order
empirical statistics, achieving the empirical capacity
corresponding to more elaborate models.

\appendix

\section{Proofs of Lemmas}\label{app:lemmas}

\begin{proof}[Proof of Lemma \ref{lem:entopry_Linf_bound}]
Let $\vect{p} = \left(p_0,p_1,\ldots,p_{|\sst \m{X}|-1}\right)$,
and assume without loss of generality that $\min(\vect{p}) = p_0$.
For any $i\in\brk{|\m{X}|}$, define
\begin{equation*}
\vect{p}^{(i)} =
\big{(}\underbrace{0,0,\ldots,0}_i,1,\underbrace{0,\ldots,0}_{|\m{X}|-i-1}\big{)}
\end{equation*}
and let $\vect{p}_u$ be the uniform distribution over $\m{X}$.
Express $\vect{p}$ as the following convex combination:
\begin{equation*}
\vect{p} = p_0\cdot|\m{X}|\cdot\vect{p}_u + \sum_{i=1}^{|\m{X}|-1}
\left(p_i-p_0\right)\cdot\vect{p}^{(i)}
\end{equation*}
Using Jensen's inequality we get
\begin{equation}\label{eq:entropy_bound_by_min}
H(\vect{p}) \geq  p_0\cdot|\m{X}|\cdot H\left(\vect{p}_u\right)
+\sum_{i=1}^{|\m{X}|-1} \left(p_i-p_0\right)\cdot
H(\vect{p}^{(i)}) = p_0\cdot|\m{X}|\log{|\m{X}|}
\end{equation}
Now, by definition we have $\|\vect{p}-\vect{p}_u\|_\infty\geq
\frac{1}{|\m{X}|}-p_0$, and hence $p_0\geq
\frac{1}{|\m{X}|}-\|\vect{p}-\vect{p}_u\|_\infty$. Substituting
this into (\ref{eq:entropy_bound_by_min}), we obtain
\begin{equation*}
H(\vect{p}) \;\geq\; \log{|\m{X}|}\Big{(}1 -
|\m{X}|\big{\|}\vect{p}-\vect{p}_u\big{\|}_\infty\Big{)}
\end{equation*}
as desired. Note that (\ref{eq:entropy_bound_by_min}) is in fact a
uniformly better bound, but the $\m{L}_\infty$ bound is sufficient
for our needs and easier to work with.

\end{proof}

\begin{proof}[Proof of Lemma \ref{lem:KTb}:]
Let $\ell_k^{i}$ be the position within the sequence $z^n$ of the
$k$th appearance of the $i$th symbol, and denote $d=d(z^n,w^n)$.
We bound the excess redundancy incurred by using the KT($b$)
estimator with noisy observations, assuming that $b+d\leq n+1$, as
follows: 
\begin{align*}
-\log{\frac{\widehat{p}^{\;\sst{KT(b)}}(z^n\|w^n)}{\widehat{p}^{\;\sst{KT}}(z^n)}}
\;&=\; \sum_{i\in\m{X}}\sum_{k=1}^{n_i(z^n)}
\log\frac{\widehat{p}^{\,\sst{KT}}(i|z^{\ell_k^{i}-1})}{\widehat{p}^{\,\sst{KT}}(i|w^{\nu_{\ell_k^{i}}})}
= \sum_{i\in\m{X}}\sum_{k=1}^{n_i(z^n)}
\log{\left(\frac{\nu_{\ell_k^{i}}+\frac{|\m{X}|}{2}}{n_i(w^{\nu_{\ell_k^{i}}})+\frac{1}{2}}\cdot
\frac{(k-1)+\frac{1}{2}}{(\ell_k^{i}-1)+\frac{|\m{X}|}{2}}\right)}
\\
&\;\leq\; \sum_{i\in\m{X}}\sum_{k=1}^{n_i(z^n)}\log\left
(\frac{\ell_k^{i}-1+\frac{|\m{X}|}{2}}{\left|k-b-d\right|^+ +
\frac{1}{2}}\cdot
\frac{k-\frac{1}{2}}{\ell_k^{i}-1+\frac{|\m{X}|}{2}}\right ) \leq
|\m{X}|\sum_{k=1}^{n}\log\left
(\frac{k-\frac{1}{2}}{\left|k-b-d\right|^+ + \frac{1}{2}}\right)
\\
\;&\leq\; |\m{X}|(b+d-1)\log(2(b+d-1)) +
|\m{X}|\sum_{k=b+d}^{n}\log\left
(1+\frac{b+d-1}{k-b-d+\frac{1}{2}}\right)
\\
&\;\leq\; |\m{X}|(b+d-1)\log(2(b+d-1)) +
|\m{X}|(b+d-1)\log{e}\sum_{k=0}^{n-1}\frac{1}{k+\frac{1}{2}}
\\
&\;\leq\; |\m{X}|(b+d-1)\log(2(b+d-1)) +
|\m{X}|(b+d-1)(2\log{e}+\log{(2n-1)})
\\
&\;\leq\; 2|\m{X}|(b+d-1)\log{2ne} \label{eq:thrm41}
\end{align*}

This completes the proof for $b+d\leq n+1$. For $b+d>n+1$, we note
that the maximal possible excess redundancy per symbol at each
step is upper bounded by $\log(2n+|\m{X}|-2)$, hence the total
excess redundancy is upper bounded by
\begin{equation*}
n\log(2n+|\m{X}|-2)\leq (b+d-1)\log(2n+|\m{X}|-2) \leq
2|\m{X}|(b+d-1)\log{2ne}
\end{equation*}
where the last transition is a simple easy exercise using the fact
that $|\m{X}|\geq 2$.
\end{proof}

\begin{proof}[Proof of Lemma \ref{lem:hoeffding_sampling}:]
We first prove the Lemma for $\m{X}=\{0,1\}$ and a deterministic
$Z^n=z^n$, the general case then results as a simple corollary.
Under these assumptions, we have
\begin{align}\label{eq:finite_sample_bound}
&\Prob\left(\,\left\|\vect{p}_{\rm emp}(z^n)-\vect{p}_{\rm
emp}\left(z^n\samp B^n\right)\right\|_\infty>\tau\right) =
\Prob\left(\,\left|p_{\rm emp}(z^n)-p_{\rm emp}\left(z^n\samp
B^n\right)\right|>\tau\right)
\\
\nonumber &=
\Prob\left(\,\left|\frac{1}{n}\sum_{k=1}^nz_k-\frac{1}{m}\sum_{k=1}^mz_{\sigma_k(B^n)}\right|>\tau\right)
\end{align}
Hoeffding showed that the distribution of the sample mean for
sampling without replacement from a finite (deterministic)
population, obeys the same bounds for deviation from the mean as
the ones he obtained for an i.i.d. sample with mean equal to the
empirical mean of the population \cite[sec. 6]{hoeffding}.
Following this, let $A^m$ be an i.i.d. $\sim{\rm
Ber}\left(\frac{1}{n}\sum_{k=1}^nz_k\right)$ sequence. Using the
standard Hoeffding inequality \cite[sec. 2]{hoeffding}, we obtain
\begin{equation*}
\Prob\left(\,\left|\frac{1}{n}\sum_{k=1}^nz_k -
\frac{1}{m}\sum_{k=1}^mA_k \right|>\tau\right) \leq
2\exp\left(-2m\tau^2\right)
\end{equation*}
and by Hoeffding's claim above, this bound also holds for
(\ref{eq:finite_sample_bound}). This concludes the proof for the
binary alphabet\footnote{In the binary case we get a coefficient
of $2$ instead of $4$ multiplying the exponent.} with a
deterministic $Z^n$. For a stochastic $Z^n$ the result stands
since $B^n$ and $Z^n$ are independent, and the bound holds for any
realization $Z^n=z^n$. For a larger alphabet, one can define an
indicator sequence $Z^n_{(i)}$ for any symbol $i\in\m{X}$, namely
a sequence whose $k$th element is given by
$Z_{k,(i)}\dfn\ind_i(Z_k)$. The proof then follows from the binary
alphabet analysis, and the union bound over all the symbols.
\end{proof}

\begin{proof}[Proof of Lemma \ref{lem:azuma_sampling}:]
For any $i\in\m{X}$, define the r.v. sequence $A^n_{(i)}$ of
length $n$, whose $k$th element is given by
\begin{equation*}
A_{k,(i)}=\sum_{j=1}^k\ind_i(Z_j)-q^{-1}\ind_i(Z_j)\cdot B_j
\end{equation*}
We have that
\begin{align*}
\Expt(A_{k+1,(i)}\big{|}A^k_{(i)}=a^k) &=  a_k +
\Expt(\ind_i(Z_{k+1})\cdot(1-q^{-1}B_{k+1})\big{|}A^k_{(i)}=a^k) \\
&=a_k +
\left[1-q^{-1}\Expt(B_{k+1}\big{|}A^k_{(i)}=a^k)\right]\Expt(\ind_i(Z_{k+1})\big{|}A^k_{(i)}=a^k)
\\
&= a_k
\end{align*}
where in the last two transitions we used the causal independence
assumption the fact that $B_k\sim{\rm Ber}(q)$. Similarly, we also
have that $\Expt(A_{1,(i)})=0$. Therefore for any $i\in\m{X}$ ,
$A^n_{(i)}$ is a zero mean martingale with differences that are
bounded by $|A_{k+1,(i)}-A_{k,(i)}|\leq \max\{1,q^{-1}-1\}\leq
q^{-1}$. By the Azuma-Hoeffding inequality for bounded-difference
martingales \cite{Chung2006}, for any $\tau>0$
\begin{equation*}
\Prob\big{(}\,\left|A_{n,(i)}\right|>n\tau\big{)} \leq
2\exp\left(-\frac{n\tau^2q^2}{2}\right)
\end{equation*}
The result is now established as follows:
\begin{align*}
\Prob&\left(\,\left\|\vect{p}_{\rm
emp}(Z^n)-\alpha(B^n)\vect{p}_{\rm emp}\left(Z^n\samp
B^n\right)\right\|_\infty>\tau\right) =
\Prob\left(\,\bigcup_{i\in\m{X}}\left\{\bigg{|}\frac{1}{n}\sum_{k=1}^n\ind_i(Z_k)-\frac{1}{nq}\sum_{k=1}^n\ind_i(Z_k)\cdot
B_k\bigg{|}>\tau\right\}\right)
\\
& \;\leq\;
\sum_{i\in\m{X}}\Prob\big{(}\,\left|A_{n,(i)}\right|>n\tau\big{)}
\leq 2|\m{X}|\exp\left(-\frac{n\tau^2q^2}{2}\right)
\end{align*}
\end{proof}

\begin{proof}[Proof of Lemma \ref{lem:sampling}:]
To avoid heavy indexing we prove the result for the first block
and the training pattern sequence, this is easily extended to any
other block and to the repetition pattern sequences. For $1\leq
k\leq b$, let $A_k$ be a r.v. denoting the exact usage of each
position, thus taking values over a generic alphabet
$\m{A}=\{active,regular,training\}\cup\brk{(|\m{X}|-1)s}$ where
the numerical values correspond to update positions for a specific
update bit using a specific input pair, as described in subsection
\ref{subsec:modulo-additive}. For brevity, we omit subscripts and
write $\Prob(x|y)$ for $\Prob_{X|Y}(x|y)$. Summations are
understood to be taken over all feasible values of the summation
variables. Let us now prove that $A_{k+1}$ and $(Z^{k+1},A^k)$ are
statistically independent for $b_a\leq k <b$, from which the
result then follows immediately, since the training pattern is
simply an indicator sequence $B_k=\ind_{training}(A_{k+b_a})$. We
have,
\begin{equation*}
\Prob(a_{k+1}|z^{k+1},a^k) =
\frac{\Prob(a^{k+1})\Prob(z^{k+1}|a^{k+1})}{\Prob(a^k)\Prob(z^{k+1}|a^k)}
= \Prob(a_{k+1}) \frac{\Prob(z^{k+1}|a^{k+1})}{\Prob(z^{k+1}|a^k)}
\end{equation*}
where in the second transition we used the fact that by
construction, $A^{b_a}$ is a constant deterministic sequence and
$A_{b_a+1}^b$ is an i.i.d. sequence. It is therefore sufficient to
show that $Z^{k+1}\leftrightarrow A^k\leftrightarrow A_{k+1}$. To
that end: {\allowdisplaybreaks
\begin{align}\label{eq:Z-A-A}
\nonumber\Prob(z^{k+1}|a^{k+1})&= \sum_{x^ky^k}
\Prob(z^{k+1}|x^k,y^k,a^{k+1})\Prob(x^k,y^k|a^{k+1})
\\\nonumber
&=\sum_{x^ky^k}
\Prob(z^k|x^k,y^k,a^{k+1})\Prob(z_{k+1}|z^k,x^k,y^k,a^{k+1})\Prob(x^k,y^k|a^{k+1})
\\\nonumber
&=\sum_{x^ky^k}\Prob(z^k|x^k,y^k)\Prob(z_{k+1}|x^k,y^k,a^{k+1})\Prob(x^k,y^k|a^{k+1})
\\\nonumber
&\stackrel{\rm
(\ref{fact1})}{=}\sum_{x^ky^k}\Prob(z^k|x^k,y^k)\Prob(z_{k+1}|x^k,y^k,a^k)\Prob(x^k|y^k,a^{k+1})\Prob(y^k|a^{k+1})
\\\nonumber
&\stackrel{\rm
(\ref{fact2})}{=}\sum_{x^ky^k}\Prob(z^k|x^k,y^k)\Prob(z_{k+1}|x^k,y^k,a^k)\Prob(x^k|y^{k-1},a^k)\Prob(y^k|a^{k+1})
\\\nonumber
&=\sum_{x^ky^k}\Prob(z^k|x^k,y^k)\Prob(z_{k+1}|x^k,y^k,a^k)\Prob(x^k|y^{k-1},a^k)\prod_{j=1}^k\Prob(y_j|y^{j-1},a^{k+1})
\\\nonumber
&\stackrel{\rm
(\ref{fact3})}{=}\sum_{x^ky^k}\Prob(z^k|x^k,y^k)\Prob(z_{k+1}|x^k,y^k,a^k)\Prob(x^k|y^{k-1},a^k)\prod_{j=1}^k\Prob(y_j|y^{j-1},a^j)
\\
&\stackrel{\rm (\ref{fact4})}{=} \Prob(z^{k+1}|a^k)
\end{align}}
Where the transitions are justified as follows ($b_a<k\leq b$):
\begin{enumerate}[(a)]

\item\label{fact1} $Z_k\leftrightarrow X^{k-1}Y^{k-1}
\leftrightarrow A^k$.

Proof: We easily find that $Z_k\leftrightarrow X^{k-1}Y^{k-1}
\leftrightarrow U^{k-1}$ by combining $Z_k\leftrightarrow
X^{k-1}Y^{k-1} \leftrightarrow X_k$ given in
(\ref{eq:Markov_relation_mod_add}), with $Z_k\leftrightarrow
X^kY^{k-1} \leftrightarrow U^{k-1}$ which is equivalent to
(\ref{eq:markov_relation}). The relation now follows since by
construction $A^k={\rm func}(U^{b_a})$.

\item\label{fact2} $X^k={\rm func}(\theta_0,Y^{k-1},A^k)$, by
construction.

\item $Y_k\leftrightarrow Y^{k-1}A^k\leftrightarrow
A_{k+1}^b$.\label{fact3}

Proof:
\begin{align*}
\Prob(y_k|y^{k-1},a^b) &= \sum_{x^k}
\Prob(y_k|y^{k-1},x^k,a^b)\Prob(x^k|y^{k-1},a^b) = \sum_{x^k}
\Prob(y_k|y^{k-1},x^k,a^k)\Prob(x^k|y^{k-1},a^k)
\\
&=\Prob(y_k|y^{k-1},a^k)
\end{align*}
where in the second transition we used (\ref{fact2}) above
together with $Y_k\leftrightarrow X^kY^{k-1} \leftrightarrow A^b$,
which stems from (\ref{eq:markov_relation}) using the fact that by
construction $A^b={\rm func}(U^{b_a})$.

\item The dependence of the expression on $a_{k+1}$ has been
removed. \label{fact4}

\end{enumerate}

\end{proof}

\begin{proof}[Proof of Lemma \ref{lem:derivative}]
Let $\vect{p},\vect{q}$ be two distributions over $\m{X}$. Then
for any $a\in\RealF^+$ such that $\vect{p} +a(\vect{q}-\vect{p})$
is also a probability distribution,
\begin{align*}
\frac{\partial}{\partial a}H\big{(}\vect{p}
+a(\vect{q}-\vect{p})\big{)} &=
-\sum_{i\in\m{X}}(q_i-p_i)\log(p_i+a(q_i-p_i)) -
\sum_{i\in\m{X}}(q_i-p_i)\frac{p_i+a(q_i-p_i)}{p_i+a(q_i-p_i)}\log{e}
\\
&= -\sum_{i\in\m{X}}(q_i-p_i)\log(p_i+a(q_i-p_i))
\\
&=
-\frac{1}{a}\left(\sum_{i\in\m{X}}(p_i+a(q_i-p_i))\log(p_i+a(q_i-p_i))
- \sum_{i\in\m{X}}p_i\log(p_i+a(q_i-p_i))\right)
\\
&= \frac{1}{a}\Big{(}H\big{(}\vect{p}
+a(\vect{q}-\vect{p})\big{)}-H(\vect{p})-D(\vect{p}\,\|\,\vect{p}
+a(\vect{q}-\vect{p}))\Big{)}
\end{align*}
Using the above, we have that for any $\beta\in [0,1]$ such that
$\vect{p} +\beta^{-1}(\vect{q}-\vect{p})$ is a probability
distribution,
\begin{align*}
\frac{\partial}{\partial
\beta}\Big{\{}&\beta\left(\log{|\m{X}|}-H\big{(}\vect{p}
+\beta^{-1}(\vect{q}-\vect{p})\big{)}\right)\Big{\}}=
\\
&= \log{|\m{X}|}-H\big{(}\vect{p}
+\beta^{-1}(\vect{q}-\vect{p})\big{)}+\Big{(}H\big{(}\vect{p}
+\beta^{-1}(\vect{q}-\vect{p})\big{)}-H(\vect{p})-D(\vect{p}\,\|\,\vect{p}
+\beta^{-1}(\vect{q}-\vect{p}))\Big{)}
\\
&= \log{|\m{X}|} -H(\vect{p}) - D(\vect{p}\,\|\,\vect{p}
+\beta^{-1}(\vect{q}-\vect{p}))
\\
&\;\leq\; \log{|\m{X}|} -H(\vect{p}) - \beta
D(\vect{p}\,\|\,\vect{p} +\beta^{-1}(\vect{q}-\vect{p})) +
(1-\beta) D(\vect{p}\,\|\vect{p})
\\
&\;\stackrel{(\rm a)}{\leq}\; \log{|\m{X}|} -H(\vect{p}) -
D(\vect{p}\,\|\,\beta(\vect{p}
+\beta^{-1}(\vect{q}-\vect{p}))+(1-\beta)\vect{p}) \;\leq\;
\log{|\m{X}|} -H(\vect{p}) - D(\vect{p}\,\|\vect{q})
\end{align*}
where in (a) we have used the convexity of the relative entropy.
The proof is concluded by substituting $\vect{p},\vect{q}$ with
$\vect{p}^{\rm reg}_{d},\vect{p}^{\rm reg}$ respectively.
\end{proof}

\section{A Horizon Free Universal Scheme}\label{app:horizon_free}
In this subsection we show how the presented finite-horizon
feedback transmission scheme can be transformed into a
horizon-free scheme, with an instantaneous rate approaching the
empirical capacity. To motivate this generalization, suppose one
wishes to transmit a fixed number of bits using the finite-horizon
scheme. In this case, it may be that capacity-wise, the receiver
could have potentially decoded enough bits half way throughout
transmission, and even worse - could not do so when transmission
ends due to a deterioration in channel conditions. In this case it
is therefore critical that the transmission can be stopped at any
given time, while achieving the instantaneous empirical capacity.

The idea is that instead of taking a fixed transmission period $n$
and dividing it into blocks of a fixed size $b(n)$, a variable
block size is used, growing with time. The apparent difficulty
with this approach is that, in contrast to the finite-horizon case
and although the size of the last block is increasing, the size of
any specific block is constant, and a non-negligible update
decoding error probability in each specific block is incurred.
This in turn results in two problems. First, bounding the error
probability as before using a union bound over update decoding
error events, provides a non-vanishing bound dominated by the
first block. Second, the resulting KT estimates use noisy
observations, which incurs a redundancy penalty. The first problem
is essentially solved by making sure that in the event where the
last accepted block is not ``recent enough'', no bits are decoded.
Loosely speaking, this event implies the empirical capacity is
small anyway with high probability, hence the resulting excess
redundancy is negligible. As for the second problem, for a
suitable selection of scheme parameters we can show that with high
probability, the hamming distance between the noise sequence and
the corresponding noisy observations sequence increases slowly
enough, so that the excess redundancy becomes negligible.

Following this discussion, the horizon-free scheme is obtained via
the following modifications of the finite-horizon scheme:
\begin{enumerate}[(A)]
\item The size of the $k$th block is set to $b_k$, where
$\{b_k\in\NaturalF\}_{k=1}^\infty$ is strictly increasing. In our
proof, we use an arithmetically growing block size\footnote{Other
block size increments are possible, resulting in different
trade-offs between error probability and convergence rate. For
instance, one can use the recursion $b_k = (\sum_{j=1}^{k-1}
b_j)^\nu$ for some $\nu \in(0,1)$.}, i.e., $b_k=b_0+k$ for some
$b_0\in\NaturalF$.

\item The parameters of the $k$th block are fixed functions of its
size $b_k$, i.e., $m_k=m(b_k), \tau_k=\tau(b_k)$, etc.

\item On the $k$th block, the update information consists of the
type of the noise sequence (symbols occurrences vector) over
regular positions in the previously accepted block, and the index
of the corresponding message interval. Let $n_k\dfn
\sum_{j=1}^{k}b_j$ be the number of channel uses in the first $k$
blocks. The number of uncoded update bits in the $k$th block is
therefore given by replacing $b\rightarrow b_{k-1}\,,n\rightarrow
n_{k-1}$ in the left-hand side of
(\ref{eq:update_bits_num_nonbinary}).

\item Transmission can be terminated at any point. When
terminated, the receiver normally decodes the binary interval
pertaining to the last known message interval (from the last
accepted block) using the corresponding ambiguity resolving bit.
However, if transmission ended during the $k$th block and the last
accepted block $k_{\,\rm acc}$ is not recent enough, namely
$k_{\,\rm acc}<\rho_k$ for some predetermined \textit{recency
threshold} $\rho_k$, then the decoded interval is $[0,1)$, i.e.,
no bits are decoded.
\end{enumerate}

We now turn to prove that this modified scheme achieves the
empirical capacity for a suitable selection of parameters
$b_k,m_k,\tau_k,\rho_k$, where the thresholds $\tau_{{\sst d},k}$
and $\tau_{u,k}$ are determined by $\tau_k$ as before. For
simplicity of the exposition, assume the length of the $k$th block
is $b_k = k$, and choose $m_k$ and $\tau_k$ to be
\begin{equation*}
m_k \dfn b_k^{a_1} = k^{a_1} \,,\quad \tau_k \dfn b_k^{-a_2} =
k^{-a_2}
\end{equation*}
for some constants $a_1,a_2\in(0,1)$. For brevity, we mostly
disregard non-integer issues throughout this section, as these
have no asymptotic effect. Let $n_k=\sum_{j=1}^k b_j =
\frac{k(k+1)}{2}$. The number of uncoded update bits in the $k$th
block is zero padded up to $s_k$, which is given by
\begin{align}\label{eq:upd_bit_num_horizon_free}
\left\lceil\log
b_{k-1}^{|\m{X}|-1}\right\rceil+\Big{\lceil}\log\Big{(}1+(|\m{X}|-1)n_{k-1}\Big{)}\Big{\rceil}
\leq (|\m{X}|-1)\log{k} + \log((|\m{X}|-1)k(k+1)) \leq
2|\m{X}|\log(k+1) \dfn s_k
\end{align}
Following the same derivations as in Section \ref{sec:analysis},
we have that
\begin{align*}
&\Prob(E_1^{(k)}) \leq 2|\m{X}|\exp\left(-\frac{1}{2}\tau_k^2m_k^2b_k^{-1}\right) =  2|\m{X}|\exp\left(-\frac{1}{2}k^{\,2(a_1-a_2)-1}\right)\\
&\Prob(E_{2}^{(k)}) \leq
4|\m{X}|^3\left\lceil\log(k+1)\right\rceil\exp\left(-\frac{\tau_k^2m_k^2b_k^{-1}}{8|\m{X}|^4\left\lceil\log(k+1)\right\rceil^2}\right)
=
4|\m{X}|^3\left\lceil\log(k+1)\right\rceil\exp\left(-\frac{k^{2(a_1-a_2)-1}}{8|\m{X}|^4\left\lceil\log(k+1)\right\rceil^2}\right)
\end{align*}
where $E_1^{(k)},E_2^{(k)}$ are the events in the $k$th block
corresponding to $E_1,E_2$ defined in subsection
\ref{subsec:err_prob}. Thus, $E_1^{(k)}\cup E_2^{(k)}$ constitutes
a necessary condition for an update decoding error in the $k$th
block.
At this stage in the finite-horizon proof, we have used the union
bound over update decoding error events in each block to obtain an
upper bound for the error probability in the finite-horizon
scheme. However, in this case taking the union bound would result
in an error probability that is dominated by the first block,
hence not decaying to zero. From this point, assume the
transmission scheme was terminated at time $n=n_k$ after
\textit{precisely $k$ arithmetically growing blocks were
sent}\footnote{In general, the scheme may be terminated
arbitrarily in the middle of some block, and decoding will be
carried out w.r.t. the end of the previous block. Since the
maximal block size is $\bigo(\sqrt{n})$, this can cause a maximal
rate fluctuation of $\bigo(n^{-\frac{1}{2}})$, which should be
added to the redundancy term $\eps_2(n)$, but turns out to be
asymptotically negligible.}, which means that $\sqrt{2n}-1\leq
k\leq \sqrt{2n}$. Let us divide the transmission period into two
\textit{batches}: The \textit{first batch} includes the first
$k^{a_3}$ blocks for some $a_3\in (0,1)$, while the \textit{last
batch} includes all the rest $k-k^{a_3}$ blocks. Let us also set
the \textit{recency threshold} to be $\rho_k \dfn k^{a_3}$, which
means that if the last accepted block resides in the first batch,
no bits are decoded.

Define $V_0$ to be the following event:
\begin{equation*}
V_0 \;\dfn\; \bigcap_{j=\lceil\rho_k\rceil}^k
\left\{\left(E_1^{(j)}\right)^c\cap
\left(E_2^{(j)}\right)^c\right\}
\end{equation*}
Using the same ideas as in the fixed-horizon analysis, we can show
$V_0$ implies that no update decoding error occurred in the last
batch. Due to the recency threshold, this implies in turn that
either the decoded message interval is correct, or that no bits
are decoded. Therefore, $V_0^c$ is a necessary condition for an
error, and so
\begin{align}\label{eq:err_prob_nonbinary_horizon_free_bound}
\nonumber
&\hspace{-0.13cm}\sup_{\begin{tiny}\begin{array}{cc}\m{W}\in\mathscr{M}_{\sstm{X}}\\\theta_0\in[0,1)\end{array}\end{tiny}}\hspace{-0.3cm}p_e(n,\m{W},\theta_0)
\leq \Prob(V_0^c) \leq k\cdot\Prob\left(E_1^{(k^{a_3})}\cup
E_2^{(k^{a_3})}\right)\leq 5|\m{X}|^3k\left\lceil\log
(k^{a_3}+1)\right\rceil\exp\left(-\frac{k^{\,a_3(2(a_1-a_2)-1)}}{8|\m{X}|^2\left\lceil\log(k^{a_3}+1)\right\rceil^2}\right)
\\
&\;\leq\; 5|\m{X}|^3\sqrt{2n}\lceil\log
((2n)^\frac{a_3}{2}+1)\rceil\exp\left(-\frac{(\sqrt{2n}-1)^{\,a_3(2(a_1-a_2)-1)}}{8|\m{X}|^4\lceil\log
((2n)^{\frac{a_3}{2}}+1)\rceil^2}\right) \;\dfn\;\eps_1(n)
\end{align}
where we have used the union bound over blocks in the last batch,
and the fact that the update decoding error probability of the
first block in that batch dominates the others. We get:
\begin{equation}\label{eq:err_prob_nonbinary_horizon_free}
-\log \eps_1(n) =
\Omega\left(\frac{n^{a_3(a_1-a_2-\frac{1}{2})}}{\log^2{n}}\right)
\end{equation}
so the error probability tends to zero uniformly for any selection
$a_1-a_2>\frac{1}{2}$. This concludes the error probability part
of the proof.

We now show that at any time point, the decoding rate attained by
the scheme is close to the empirical capacity with probability
approaching one. Let $V_1$ be the event where none of the blocks
in the last batch were discarded due to an improper selection of
$M_t,M_u$ made by the receiver. Using Hoeffding's inequality as in
(\ref{eq:prob_V1c}), it is readily verified that
\begin{equation}\label{eq:horizon_free_V12}
-\log\Prob(V_1^c) = \Omega\left(n^{a_3(a_1-\frac{1}{2})}\right)
\end{equation}
and so for any selection $a_1-a_2>\frac{1}{2}$ both
$\Prob(V_0),\Prob(V_1)\rightarrow 1$. Now, let $\vect{p}_a^{\rm
reg,f}$ be the empirical distribution over accepted regular
positions in the first batch, and $\vect{p}_a^{\rm reg,\ell}$ the
corresponding distribution in the last batch. Let us express
$\vect{p}_a^{\rm reg}$ as
\begin{equation*}
\vect{p}_a^{\rm reg} = \lambda\vect{p}_a^{\rm
reg,f}+(1-\lambda)\vect{p}_a^{\rm reg,\ell}
\end{equation*}

Due to possible non-negligible erroneous update decoding, the
receiver might use noisy observations for its KT estimates. In the
finite-horizon case this problem was averted since the update
error probability in \textit{each block} was negligible, and so
the event of noisy observations had a vanishing impact
incorporated into the redundancy term $\eps_3(n)$. However, in the
horizon-free case there is a non-vanishing update error
probability dominated by the first blocks. Nevertheless, under
$V_0$ only the first batch may include erroneous blocks, and thus
the \textit{hamming distance} between the actual noise sequence
(over accepted regular positions) and the one used by the receiver
when updating the KT estimates, is upper bounded by $d=\lambda n$.
Now, since $b_k\leq \sqrt{2n}$, the receiver uses a
KT$(2\sqrt{2n})$ estimator and using Lemmas \ref{lem:KT} and
\ref{lem:KTb} with noisy observations we have
\begin{equation}\label{eq:horizon_free1}
R^{\rm reg} \geq R_{\sst\beta} -
K_7\left(\frac{\log{n}}{\sqrt{n}}+\lambda\log{n}\right)
\quad\qquad \big{(}\text{\rm given }V_0\cap V_1\big{)}
\end{equation}
for some $K_7>0$ large enough, where $R_{\sst\beta} \dfn
\beta\big{(}\log{|\m{X}|}-H(\vect{p}_a^{\rm reg})\big{)}$. Let
$\xi_j$ be the maximal possible number of non-regular positions in
blocks $j$ to $k$ (where $k$ is the last block), i.e,
\begin{equation}\label{eq:xi_j_def}
\xi_j \;\dfn\;  \sum_{i=j}^k 4m_j(1+2\log{b_j}) \leq
K_3\sum_{i=j}^k m_j\log{b_j}
\end{equation}
where $K_2$ was defined in (\ref{eq:preg2pemp}). Simple algebraic
manipulations yield the following bound for $\beta$:
\begin{equation}\label{eq:beta_vs_beta_ell}
\beta \geq
\frac{\beta_\ell}{1-\lambda}\left(\frac{n-n_{\rho_k}-\xi_{\rho_k+1}}{n}\right)
=
\frac{\beta_\ell}{1-\lambda}\left(1-2n^{a_3-1}-K_8n^{\frac{a_1-1}{2}}\log{n}\right)
\end{equation}
for some $K_8>0$, where $\beta_\ell$ is the fraction of regular
positions in the last batch that were accepted. To continue, we
need the following Lemma.
\begin{lemma}\label{lem:entropy_bound2}
Let $\vect{p}$ and $\vect{q}$ be any two probability distributions
over a finite alphabet $\m{X}$. Then for any $\lambda\in[0,1]$
\begin{equation*}
H\left(\lambda\vect{q}+(1-\lambda)\vect{p}\right) \;\leq\;
H(\vect{p}) + 3|\m{X}|\lambda \log\frac{2}{\lambda}
\end{equation*}
\end{lemma}
\begin{proof}
Let $\vect{p}=(p_{\sst 1},\ldots,p_{\sst |\m{X}| })$ be a
probability distribution over $\m{X}$ with nonzero elements. Let
$\vect{v}$ be the representation of $\vect{p}$ over the
$|\m{X}|-1$ dimensional probability simplex
$\,\Simplex^{|\m{X}|-1}$, i.e., a vector of the first $|\m{X}|-1$
elements of $\vect{p}$. With some abuse of notations, denote by
$H(\vect{v})$ the entropy function of $\vect{p}$, calculated over
$\Simplex^{|\m{X}|-1}$. We take the partial derivatives of $H$ and
get
\begin{equation*}
\frac{\partial H(\vect{v})}{\partial v_i} = \log\frac{1-\sum_j
v_j}{v_i} \,,\quad i=1,\ldots,|\m{X}|-1
\end{equation*}
Since $H(\cdot)$ is concave over $\Simplex^{|\m{X}|-1}$, its
tangents at any point are always above it. Therefore for any
$\vect{r}\in\RealF^{|\m{X}|-1}$ that satisfies
$(\vect{v}+\vect{r})\in \Simplex^{|\m{X}|-1}$, we have that
\begin{equation}\label{eq:H_over_simplex}
H(\vect{v}+\vect{r}) \leq H(\vect{v}) + \sum_{i=1}^{|\m{X}|-1}r_i
\log\frac{1-\sum_j v_j}{v_i}
\end{equation}

Now let $\vect{v}$ and $\vect{u}$ be vectors over the $|\m{X}|-1$
dimensional simplex that correspond to $\vect{p}$ and $\vect{q}$
respectively, and $0\leq\lambda\leq 1$ some constant. With the
same abuse of notations, we use (\ref{eq:H_over_simplex}):
\begin{align*}
H(\lambda\vect{q}+(1-\lambda)\vect{p}) &=
H(\vect{v}+\lambda(\vect{u}-\vect{v})) \leq H(\vect{v}) +
\lambda\sum_{i=1}^{|\m{X}|-1}(u_i-v_{i}) \log\frac{1-\sum_j
v_{j}}{v_{i}} \\ &= H(\vect{p}) +
\lambda\sum_{i=1}^{|\m{X}|-1}(q_i-p_{i}) \log\frac{p_{\sst
|\m{X}|}}{p_{i}} \leq H(\vect{p}) +
\lambda\sum_{i=1}^{|\m{X}|-1}\left| \log\frac{p_{\sst
|\m{X}|}}{p_{i}}\right|
\end{align*}

Assume for the moment that $\lambda<\frac{1}{4}$. If it so happens
and all the symbol probabilities satisfy $p_i> \lambda$, then from
the above we have
\begin{equation}\label{eq:entropy_dev_bound1}
H(\lambda\vect{q}+(1-\lambda)\vect{p}) \leq H(\vect{p}) +
\lambda\sum_{i=1}^{|\m{X}|-1}\left| \log\frac{p_{\sst
|\m{X}|}}{p_{i}}\right| \leq H(\vect{p}) + (|\m{X}|-1)\lambda
\log\frac{1}{\lambda}
\end{equation}
which satisfies the statement in the Lemma. Otherwise, assume that
there are precisely $t$ symbols that do not satisfy that
requirement. Without loss of generality we assume that $p_1\leq
p_2\leq \cdots\leq p_{\sst |\m{X}|}$, and therefore $p_{\sst
t}\leq \lambda$. Define
\begin{equation*}
\vect{\psi} \dfn \lambda\vect{q}+(1-\lambda)\vect{p} = (\psi_{\sst
1},\ldots,\psi_{\sst |\m{X}|})
\end{equation*}
The first $t$ elements of $\vect{\psi}$ are all smaller than
$2\lambda$. Without loss of generality we assume that al least one
of those $t$ elements is nonzero, as otherwise we can reduce the
dimension of the problem. We have the following:
\begin{align*}
H(\vect{\psi}) \;&\stackrel{\rm (a)}{=}\;  H(\psi_{\sst
1}+\psi_{\sst |\m{X}|},\psi_{\sst 2},\ldots,\psi_{\sst |\m{X}|-1})
+ \left(\psi_{\sst 1}+\psi_{\sst |\m{X}|} \right)h_{\sst
B}\left(\frac{\psi_{\sst 1}}{\psi_{\sst 1}+\psi_{\sst
|\m{X}|}}\right)
\\
&\stackrel{\rm (b)}{\leq}  H(\psi_{\sst 1}+\psi_{\sst
|\m{X}|},\psi_{\sst 2},\ldots,\psi_{\sst |\m{X}|-1}) + h_{\sst
B}(2\lambda) \;\stackrel{\rm (c)}{\leq}\;
H\left(\sum_{i=1}^{t}\psi_{\sst i}+\psi_{\sst |\m{X}|},\psi_{\sst
t+1},\ldots,\psi_{\sst |\m{X}|-1}\right) + t\cdot h_{\sst
B}(2\lambda)
\\
&\stackrel{\rm (d)}{\leq} H\left(\sum_{i=1}^{t}p_{\sst i}+p_{\sst
|\m{X}|},p_{\sst t+1},\ldots,p_{\sst |\m{X}|-1}\right) +
(|\m{X}|-t-1)\lambda \log\frac{1}{\lambda} + t\cdot h_{\sst
B}(2\lambda)
\\
&\stackrel{\rm (e)}{\leq} H(\vect{p}) + |\m{X}|\lambda
\log\frac{1}{\lambda} +
|\m{X}|\left(2\lambda\log{\frac{1}{2\lambda}}+2\lambda\log{e}\right)
\leq H(\vect{p}) + 3|\m{X}|\lambda \log\frac{2}{\lambda}
\end{align*}
In (a) we applied the entropy's grouping property \cite{cover}, in
(b) we used the fact that $\lambda<\frac{1}{4}$, in (c) we
repeated the two previous steps $t-1$ more times, in (d) we used
(\ref{eq:entropy_dev_bound1}) since the probability vector
argument of the entropy function has a minimal symbol probability
exceeding $\lambda$ , and in (e) we used $0\leq t \leq |\m{X}|$,
the entropy's grouping property, and the inequality $h_{\sst B}(p)
\leq p\log{\frac{1}{p}} + p\log{e}$. This proves the result for
$\lambda<\frac{1}{4}$. The proof is now concluded by noticing that
for $\lambda\geq\frac{1}{4}$ the excess term satisfies
$3|\m{X}|\lambda \log\frac{2}{\lambda} > \log{|\m{X}|}$.
\end{proof}

Applying Lemma \ref{lem:entropy_bound2} to $R_{\sst\beta}$ and
using inequality (\ref{eq:beta_vs_beta_ell}), we have
\begin{align}\label{eq:horizon_free2}
\nonumber R_{\sst\beta} &\;=\;
\beta\Big{(}\log{|\m{X}|}-H(\lambda\vect{p}_a^{\rm
reg,f}+(1-\lambda)\vect{p}_a^{\rm reg,\ell})\Big{)}
\\
\nonumber&\;\geq\;
\frac{\beta_\ell}{1-\lambda}\left(1-2n^{a_3-1}-K_8n^{\frac{a_1-1}{2}}\log{n}\right)\Big{(}\log{|\m{X}|}-H(\vect{p}_a^{\rm
reg,\ell}) - 3|\m{X}|\lambda\log{\frac{2}{\lambda}}\Big{)}
\\
& \;\geq\;
\left(\frac{1-2n^{a_3-1}-K_8n^{\frac{a_1-1}{2}}\log{n}}{1-\lambda}\right)R_{\sst\beta}^\ell
- 3|\m{X}|\frac{\lambda}{1-\lambda}\log{\frac{2}{\lambda}}
\end{align}
where
\begin{equation*}
R_{\sst\beta}^\ell \dfn
\beta_\ell\Big{(}\log{|\m{X}|}-H(\vect{p}_a^{\rm reg,\ell})
\Big{)}
\end{equation*}

$R_{\sst\beta}^\ell$ is a quantity similar to $R_{\sst\beta}$, but
only for the last batch. Define $\vect{p}^\ell,\vect{p}^{{\rm
reg},\ell},\vect{p}^{{\rm reg},\ell}_d$ to be the empirical
distribution of the noise sequence in the entire last batch, over
regular positions, and over discarded regular positions,
respectively. We can now repeat the finite-horizon analysis over
the last batch only, using the parameters of the $\rho_k$-th block
which is the smallest in the batch. Namely, one can set the an
auxiliary parameter $\gamma(n)=\lito(1)$ to satisfy (the
equivalent of (\ref{eq:gamma_set}))
\begin{equation*}
\gamma^2 =
\Omega\left(m_{\rho_k}b^{-1}_{\rho_k}\log{b_{\rho_k}}\right) +
\Omega\left(\tau_{\rho_k}\right) =
\Omega\left(n^{\frac{a_1-1}{2}}\log{n}\right) + \Omega
\left(n^{-\frac{a_2}{2}}\right)
\end{equation*}
and define the events
\begin{equation}
V_2 \;\dfn\; \left\{2|\m{X}|\Linf{\vect{p}_d^{{\rm
reg},\ell}-\vect{p}_u}^{\frac{1}{2}}\leq
\frac{\gamma}{2}\right\}\,,\qquad
V_3\;\dfn\;\left\{\Linf{\vect{p}^\ell-\vect{p}_u} \geq
\gamma\right\}
\end{equation}
to obtain
\begin{equation}\label{eq:R_beta_last}
R_{\sst\beta}^\ell \geq R_{\sst\beta=1}^\ell \geq
\log{|\m{X}|}-H(\vect{p}^{{\rm reg},\ell}) \quad\qquad
\big{(}\text{\rm given }\bigcap_{i=0}^3V_i\big{)}
\end{equation}
and with
\begin{equation*}
-\log\Prob(V_2^c) =
\Omega\left(b_{\rho_k}^{-1}m_{\rho_k}^2\,\gamma^4\right) =
\Omega\left(\log^2{n}\cdot
n^{a_3(a_1-\frac{1}{2})-(1-a_1)}\right)+
\Omega\left(n^{a_3(a_1-\frac{1}{2})-a_2}\right)
\end{equation*}
and so setting $a_3(a_1-\frac{1}{2})>\max{(1-a_1,a_2)}$ we have
$\Prob(V_2)\rightarrow 1$.

Let us no define $V_4$ as the event where $\lambda<n^{-a_4}$ for
some $a_4\in(0,1)$. Using (\ref{eq:horizon_free1}),
(\ref{eq:horizon_free2}) and (\ref{eq:R_beta_last}) we
get\footnote{Notice that the minimal positive value for $\lambda$
is always greater than $\frac{1}{n}$ (single accepted block in the
first batch), and for $\lambda=0$ (first batch fully discarded)
the penalty term
$\frac{\lambda}{1-\lambda}\log{\frac{2}{\lambda}}$ is zero.}
\begin{align}\label{eq:horizon_free3}
\nonumber R^{\rm reg} &\geq \log{|\m{X}|}-H(\vect{p}^{{\rm
reg},\ell})-\frac{\left|2n^{a_3-1}+K_8n^{\frac{a_1-1}{2}}\log{n}-n^{-a_4}\right|\log{|\m{X}|}}{1-n^{-a_4}}-
3|\m{X}|\frac{n^{-a_4}\log{(2n^{a_4})}}{1-n^{-a_4}}
\\
&-K_7\left(\frac{\log{n}}{\sqrt{n}}+n^{-a_4}\log{n}\right) =
\log{|\m{X}|}-H(\vect{p}^{{\rm reg},\ell})
+\bigo\left(n^{-a_5}\log{n}\right)\qquad \big{(}\text{\rm given
}\bigcap_{i=0}^4V_i\big{)}
\end{align}
where $a_5 \dfn \min(1-a_3,\frac{1-a_1}{2},a_4,\frac{1}{2})$. To
express (\ref{eq:horizon_free3}) in terms of the rate $R_n$ and
the empirical entropy $H(\vect{p}_{\rm emp}(Z^n))$ we use Jensen's
inequality and standard manipulations, yielding (given $V_1$)
\begin{equation}\label{eq:rate_conversion}
R_n \geq \left(1-\frac{\xi_1}{n}\right)R^{\rm reg}\,,\quad
H(\vect{p}^{\ell})\geq
\left(1-\frac{\xi_{\rho_k}}{n}\right)H(\vect{p}^{{\rm reg},\ell})
\,,\quad H(\vect{p}_{\rm emp}(Z^n))\geq
\left(1-\frac{\zeta}{n}\right)H(\vect{p}^{{\rm reg},\ell})
\end{equation}
where the terms $\xi_j$ was defined in (\ref{eq:xi_j_def}) and
$\zeta$ is given by
\begin{align*}
\zeta = \xi_{\rho_k}+\sum_{j=1}^{\rho_k}b_j=
\bigo\left(n^{\frac{1+a_1}{2}}\log{n}\right) + \bigo(n^{a_3})
\end{align*}
and corresponds to the maximal number of channel uses wasted on
the first batch and on non-regular transmission in the second
batch together. Thus, since $a_5$ already involves all the
relevant terms, we get
\begin{equation}\label{eq:Rn_capVi_bound}
R_n(\m{W},\theta_0) = C^{\,\rm
emp}_n(\m{W},\theta_0)+\bigo(n^{-a_5}\log{n})\qquad
\big{(}\text{\rm given }\bigcap_{i=0}^4V_i\big{)}
\end{equation}
As before, under $V_3^c$ the $\m{L}_\infty$ bound for the entropy
(Lemma \ref{lem:entopry_Linf_bound}) yields
$\log|\m{X}|-H(\vect{p}^{\ell})\leq
|\m{X}|\log|\m{X}|\gamma(n)=\bigo\left(n^{\frac{a_1-1}{4}}\sqrt{\log{n}}\right)
+ \bigo\left(n^{-\frac{a_2}{4}}\right)$ and using
(\ref{eq:rate_conversion}) yields in turn
\begin{equation}\label{eq:V3c_free}
C^{\,\rm emp}_n(\m{W},\theta_0)=
\bigo\left(n^{\frac{a_1-1}{4}}\sqrt{\log{n}}\right) +
\bigo\left(n^{-\frac{a_2}{4}}\right)+\bigo\left(n^{\frac{a_1-1}{2}}\log{n}\right)+\bigo(n^{a_3-1})\qquad
\big{(}\text{\rm given }V_3^c\big{)}
\end{equation}
This penalty can be incorporated into the redundancy term to
remove the dependence on the event $V_3$.

We would also like to remove the dependence on $V_4$, and to that
end consider the event $V_4^c\cap V_0$. Under this event and
assuming $a_3+a_4<\frac{1}{2}$, there can be no accepted block of
size $\Omega(n^{a_3+a_4})$, as otherwise $V_4^c$ is contradicted.
Due to $V_0$, the empirical distribution (over passive positions)
of each of these larger blocks is $\tau_j$ close to the training
estimate, which in turn is $\tau_j$ close to being uniform, where
$j$ is the index/length of the corresponding block. Using the
$\m{L}_\infty$ bound for the entropy, the empirical capacity of
each of these blocks (over passive positions) is therefore
$\bigo(\tau_j)=\bigo(n^{-a_2(a_3+a_4)})$, where we have used the
fact that $\tau_j$ of the smallest such block dominates the
others. Moreover, the empirical capacity of all the blocks of size
$\bigo(n^{a_3+a_4})$ is $\bigo(1)$ (which is true of course for
any block). By convexity, the empirical capacity over the entire
transmission period is no larger than the average of the empirical
capacities over some segmentation. Hence,
\begin{equation}\label{eq:V4cV0}
C^{\,\rm emp}_n(\m{W},\theta_0)=
\bigo\left(n^{-a_2(a_3+a_4)}\right) +
\bigo\left(n^{-2(a_3+a_4)}\right)+\bigo\left(n^{\frac{1-a_1}{2}}\log{n}\right)
+ \bigo(n^{a_3-1}) \qquad \big{(}\text{\rm given }V_4^c\cap
V_0\big{)}
\end{equation}
where the first term is the contribution of blocks of size
$\Omega(n^{a_3+a_4})$, the second term is the contribution of the
smaller blocks (after averaging w.r.t. the fraction of time they
occupy), and the last two terms correspond to the deviation
possibly incurred by considering only passive positions.

Let us now combine all the above results. The following inclusion
is easily verified:
\begin{equation}
\{\bigcap_{i=0}^4 V_i\}\cup\left\{V_3^c
\right\}\cup\left\{V_4^c\cap V_0\right\} \;\supseteq\;
\bigcap_{i=0}^2 V_i
\end{equation}
Under the event on the left above (and thus also under the event
on the right) at least one of (\ref{eq:Rn_capVi_bound}),
(\ref{eq:V3c_free}) or (\ref{eq:V4cV0}) must hold. We therefore
conclude that
\begin{equation}\label{eq:horizon_free4}
\Prob\Big{(}R_n(\m{W},\theta_0) \;\geq\;  C^{\,\rm
emp}_n(\m{W},\theta_0) - \eps_2(n)\Big{)} \;\geq\;
\Prob\left(\bigcap_{i=0}^2 V_i\right) \;\geq\; 1-\eps_3(n)
\end{equation}
where all the redundancy terms are now incorporated into
$\eps_2(n)$ using all the constraints set thus far (with some
further relaxations):
\begin{equation*}
\eps_2(n) = \bigo\left(n^{-a_6}\log{n}\right)\,,\qquad a_6\dfn
\min\left(\frac{1-a_1}{4},\frac{a_2}{4},1-a_3,a_2(a_3+a_4),a_4,\frac{1}{2}\right)
\end{equation*}
and where $\eps_3(n) \leq \sum_{i=0}^2\Prob(V_i^c)$, hence
\begin{equation*}
-\log\eps_3(n)=
\Omega\left(n^{a_3(a_1-\frac{1}{2})-\max\left(1-a_1,a_2\right)}\right)
\end{equation*}

To conclude, we summarize the constraints on the constants
$a_i\in(0,1)$ which guarantee that $\eps_i(n)\rightarrow 0$, so
that the empirical capacity is achieved:
\begin{equation*}\label{eq:cond_ai_horizon_free}
\max\left(1-a_1,a_2\right)<a_3\left(a_1-\frac{1}{2}\right)\,,\quad
a_3+a_4<\frac{1}{2}
\end{equation*}
There are many parameter selections that satisfy the conditions
above, e.g. $(a_1,a_2,a_3,a_4) =
\left(\frac{7}{8},\frac{1}{8},\frac{3}{8},\frac{1}{16}\right)$.

\section{A Universal Scheme for $\m{C}_{\sstm{X}}$ Utilizing Common Randomness}\label{app:dithering}

In this section we show how the horizon-free universal scheme
developed in the previous section can be adapted, using common
randomness, to achieve the empirical capacity over the larger
family $\m{C}_{\sstm{X}}$ of all causal channels. To that end, we
first describe a more general communication setting using common
randomness, and later show how our scheme is adapted into this
setting. Note that only passive feedback of the received sequence
is assumed, due to the availability of common randomness. A
\textit{feedback transmission scheme with common randomness} is a
triplet $(G,\m{P},\Delta)$ and can operate either \textit{with or
without dithering} (the definitions and details appear below).
Using the scheme over a channel $\m{W}\in\mathscr{C}_{\sstm{X}}$
with a message point $\theta_0\in[0,1)$, is described by the
following construction:
\begin{itemize}

\item Common randomness resources are assumed to be available in
the following form:
\begin{itemize}
\item An i.i.d. \textit{control sequence} $A^\infty$ taking values
over some countable alphabet $\m{A}$, with a given sequence of
marginal distributions $\m{P}\;\dfn\;\big{\{}P_k(\cdot) =
\Prob_{A_k}(\cdot)\big{\}}_{k=1}^\infty$.

\item An i.i.d. \textit{dithering sequence} $\Phi^\infty$ taking
values over $\m{X}$. When the scheme operates with dithering then
$\Phi_k\sim{\rm Uniform}(\m{X})$, and when it operates without
dithering we set $\Phi_k=0$ for any $k\in\NaturalF$.

\end{itemize}

\item $(\wt{X}^\infty,\wt{Y}^\infty)$ constitute an input/output
pair for the channel $\m{W}\in\mathscr{C}_{\sstm{X}}$.

\item $(X^\infty,Y^\infty)$ are defined by
\begin{equation}\label{eq:dithering_def}
\wt{X}_k=X_k+\Phi_k \qquad \wt{Y}_k=Y_k+\Phi_k\,,\qquad
k\in\NaturalF
\end{equation}

\item For any message point $\theta_0\in[\,0,1)$ and any
$k\in\NaturalF$
\begin{equation}\label{eq:trans_func_dithering}
 X_k=g_k(\theta_0,Y^{k-1},A^k)
\end{equation}
where $G=\left\{g_k:[0,1)\times \m{X}^{k-1}\times \m{A}^k\mapsto
\m{X}\right\}_{k=1}^\infty$ is a sequence of \textit{transmission
functions}.

\item $A_k$ is statistically independent of
$(X^{k-1},Y^{k-1},\Phi^{k-1},A^{k-1})$ for any $k\in\NaturalF$.

\item $\Phi_k$ is statistically independent of
$(X^k,Y^{k-1},\Phi^{k-1},A^k)$ for any $k\in\NaturalF$.

\item The following Markov relation holds for any $k\in\NaturalF$:
\begin{equation}\label{eq:markov_relations_dither}
\wt{Y}_k\leftrightarrow \wt{X}^k\wt{Y}^{k-1}\leftrightarrow
A^k\Phi^k
\end{equation}
Loosely speaking, this relation guarantees privacy of randomness
resources, namely that the adversary/channel cannot utilize common
randomness shared by the terminals. This is the common randomness
counterpart of (\ref{eq:markov_relation}).

\item
$\Delta=\{\Delta_k:\m{X}^k\times\m{A}^{k-1}\mapsto\mathfrak{J}\}_{k=1}^\infty$
is a sequence of \textit{decoding rules}, such that
$\Delta_k(Y^k,A^{k-1})$ is the decoded interval at time $k$.

\end{itemize}

For any given channel $\m{W}\in\mathscr{C}_{\sstm{X}}$, feedback
transmission scheme $(G,\m{P},\Delta)$ with/without dithering and
message point $\theta_0\in[\,0,1)$, the above construction
uniquely determines the joint statistics of
$(X^\infty,Y^\infty,\linebreak\wt{X}^\infty,\wt{Y}^\infty,\Phi^\infty,A^\infty)$.
The error probability $p_e(n,\m{W},\theta_0)$ and instantaneous
rate $R_n(\m{W},\theta_0)$ are defined similarly to
(\ref{eq:rate_err_def}), with $\Delta_n(Y^n,A^{n-1})$ replacing
$\Delta_n(Y^n,U^{n-1})$. The empirical capacity $C^{\,\rm
emp}_n(\m{W},\theta_0)$ is defined as in (\ref{eq:def_emp_cap}),
using the realized noise sequence $Z^\infty$ pertaining to the
input/output pair $(\wt{X}^\infty,\wt{Y}^\infty)$. A scheme is
said to \textit{locally achieve} the empirical capacity for a
specific channel/messgae point pair $(\m{W},\theta_0)$, if
\begin{align*}
p_e(n,\m{W},\theta_0) < \eps_1(n) \,,\quad
\Prob\Big{(}R_n(\m{W},\theta_0)>C^{\,\rm
emp}_n(\m{W},\theta_0)-\eps_2(n)\Big{)}
> 1-\eps_3(n)
\end{align*}
for some $\eps_1(n),\eps_1(n),\eps_1(n)\rightarrow 0$. As in
(\ref{eq:achv_emp_cap_def}), the scheme is said to (uniformly)
achieve the empirical capacity over a family of channels $\m{F}$,
if the above is satisfied uniformly over $\m{W}\in\m{F}$ and
$\theta_0\in[0,1)$.

Assuming the scheme operates with dithering, let
$\m{W}^\dagger\in\mathscr{C}_{\sstm{X}}$ denote the causal channel
induced by the input/output pair $(X^\infty,Y^\infty)$, i.e., the
channel defined by
\begin{equation}\label{eq:induced_channel}
\m{W}^\dagger(G,\m{P},\Delta,\m{W},\theta_0)\;\dfn\;
\left\{W^\dagger_k(y_k|x^k,y^{k-1})=
\Prob_{Y_k|X^kY^{k-1}}(y_k|x^k,y^{k-1})\right\}_{k=1}^\infty
\end{equation}
The \textit{induced channel} $\m{W}^\dagger$ depends in general on
the transmission scheme and the message point, and is therefore
not a ``true channel'' in the regular operational sense. Moreover,
generally $\m{W}^\dagger\not\in\mathscr{M}_{\sstm{X}}$ despite the
modulo-additive dithering, due to the statistical coupling
generated by feedback\footnote{Note however that in the special
case where $\m{W}$ is memoryless, the induced channel
$\m{W}^\dagger$ is independent of the transmission scheme and the
message point, is memoryless and modulo-additive, and is obtained
by averaging ``cyclicly shifted'' versions of $\m{W}$ (see the
discussion in the end of section \ref{sec:result} for the binary
case).}. Nevertheless, the following observation provides an
operational meaning to the induced channel.
\begin{lemma}\label{lem:equivalence}
Fix a channel $\m{W}\in\mathscr{C}_{\sstm{X}}$ and a message point
$\theta_0\in[0,1)$. Let
$\m{W}^\dagger=\m{W}^\dagger(G,\m{P},\Delta,\m{W},\theta_0)$ be
the corresponding induced channel. The following two statements
are equivalent:
\begin{enumerate}[(i)]
\item \label{case1} The scheme $(G,\m{P},\Delta)$ operating
\textit{with dithering} locally achieves the empirical capacity
for $(\m{W},\theta_0)$, with the convergence parameters
$\eps_1(n),\eps_2(n),\eps_3(n)$.

\item \label{case2} The scheme $(G,\m{P},\Delta)$ operating
\textit{without dithering} locally achieves the empirical capacity
for $(\m{W}^\dagger,\theta_0)$, with the convergence parameters
$\eps_1(n),\eps_2(n),\eps_3(n)$.
\end{enumerate}
\end{lemma}
\begin{proof}
For any feedback transmission scheme operating with or without
dithering, the decoded interval $\Delta_n$ is a function of
$(Y^n,A^{n-1})$, and the rate and error probability are in turn
functions of $\Delta_n$. Moreover, the realized noise sequence
corresponding to $(\wt{X}^\infty,\wt{Y}^\infty)$ and to
$(X^\infty,Y^\infty)$ is exactly the same sequence due to
(\ref{eq:dithering_def}), hence the empirical capacity $C^{\,\rm
emp}_n(\m{W},\theta_0)$ is a function of $(X^n,Y^n)$. Therefore in
general, $(\Delta_n,p_e,R_n,C^{\,\rm emp}_n)$ are functions of
$(X^n,Y^n,A^{n-1})$. Now for case (\ref{case2}) above, the induced
channel $\m{W}^\dagger$ together with $\theta_0$ and
$(G,\m{P},\Delta)$ uniquely defines the joint distribution of
$(X^\infty,Y^\infty,A^\infty)$. But by definition, this
distribution must coincide with the joint distribution of
$(X^\infty,Y^\infty,A^\infty)$ obtained in case (\ref{case1}),
concluding the proof.
\end{proof}

It should be emphasized that the two statements in the Lemma above
correspond to two separate constructions. The following important
observation is due.
\begin{lemma}\label{lem:sampling_dithering}
Let $\m{W}\in\mathscr{C}_{\sstm{X}}$, and suppose the scheme
$(G,\m{P},\Delta)$ operates without dithering over the channel
$\m{W}^\dagger(G,\m{P},\Delta,\m{W},\theta_0)$ with the message
point $\theta_0$. Then for any $a\in\m{A}$, the indicator sequence
$\left\{\ind_a(A_k)\right\}_{k=1}^\infty$ is a (not necessarily
identically distributed) causal sampling sequence for the noise
sequence $Z^\infty$.
\end{lemma}
\begin{proof}
This is an analogue to the statement made in Lemma
\ref{lem:sampling}, and the proof is of the same spirit. We will
prove for the case where $(G,\m{P},\Delta)$ operates with
dithering over $\m{W}$ with $\theta_0$, and the result will follow
as in Lemma \ref{lem:equivalence}, since the distribution of
$(Z^\infty,A^\infty)$ under both settings is the same. Clearly,
$\ind_a(A_k)\sim{\rm Ber}(P_k(a))$ and the indicator sequence is
not necessarily identically distributed. However as we now show,
$A_k$ is statistically independent of $(Z^k,A^{k-1})$ for any
$k\in\NaturalF$, from which the result follows immediately. Since
$A^\infty$ is a sequence of independent r.v's it is sufficient to
show that $Z^k\leftrightarrow A^{k-1} \leftrightarrow A_k$. To
this end, we repeat the derivation in (\ref{eq:Z-A-A}) to the
letter, replacing transition justifications (\ref{fact1}) and
(\ref{fact3}) with the following (\ref{factD1}${}^*$) and
(\ref{factD2}${}^*$) respectively:
\begin{enumerate}[(a*)]

\item $Z_k\leftrightarrow X^{k-1}Y^{k-1}A^{k-1} \leftrightarrow
A_k$.\label{factD1}

Proof: Again we omit the r.v. subscripts where there is no
confusion, vector additions over $\m{X}^k$ are taken to be element
by element modulo-addition.{\allowdisplaybreaks
\begin{align*}
&\Prob(z_k|x^{k-1},y^{k-1},a^k) = \sum_{\phi^k}
\Prob(\phi_k|\phi^{k-1},x^{k-1},y^{k-1},a^k)\Prob(\phi^{k-1}|x^{k-1},y^{k-1},a^k)\Prob(z_k|x^{k-1},y^{k-1},a^k,\phi^k)
\\
&\stackrel{(\rm \ref{fact_a1})}{=}\sum_{\phi^k}\Big{[}
\Prob(\phi_k)\Prob(\phi^{k-1}|x^{k-1},y^{k-1},a^k)
\\
&\times
\Prob_{\wt{Y}_k|\wt{X}^k,\wt{Y}^{k-1}A^k\Phi^k}\Big{(}g_k(\theta_0,y^{k-1},a^k)+\phi_k+z_k\big{|}g_k(\theta_0,y^{k-1},a^k)+\phi_k,x^{k-1}+\phi^{k-1},y^{k-1}+\phi^{k-1},a^k,\phi^k\Big{)}\Big{]}
\\
&\stackrel{(\rm \ref{fact_a2})}{=}|\m{X}|^{-1}\sum_{\phi^{k-1}}
\Big{[}\Prob(\phi^{k-1}|x^{k-1},y^{k-1},a^{k-1})
\\
&\qquad\qquad \times
\sum_{\phi_k\in\m{X}}W_k\left(g_k(\theta_0,y^{k-1},a^k)+\phi_k+z_k|g_k(\theta_0,y^{k-1},a^k)+\phi_k,x^{k-1}+\phi^{k-1},y^{k-1}+\phi^{k-1}\right)\Big{]}
\\
&\stackrel{(\rm \ref{fact_a3})}{=}|\m{X}|^{-1}\sum_{\phi^{k-1}}
\Prob(\phi^{k-1}|x^{k-1},y^{k-1},a^{k-1})\sum_{\phi^{\,'}_k\in\m{X}}W_k\left(\phi^{\,'}_k+z_k|\phi^{\,'}_k,x^{k-1}+\phi^{k-1},y^{k-1}+\phi^{k-1}\right)
\\
&\stackrel{(\rm
\ref{fact_a4})}{=}\Prob(z_k|x^{k-1},y^{k-1},a^{k-1})
\end{align*}}
where transitions are justified as follows:
\begin{enumerate}[({a}1)]
\item \label{fact_a1} $\Phi_k$ is statistically independent of
$(X^{k-1},Y^{k-1},\Phi^{k-1},A^k)$, together with
(\ref{eq:dithering_def}) and (\ref{eq:trans_func_dithering}).

\item \label{fact_a2} $\Phi_k$ is uniformly distributed, $A_k$ is
statistically independent of $(X^{k-1},Y^{k-1},\Phi^{k-1})$, the
Markov relation (\ref{eq:markov_relations_dither}) and the
definition of the channel $\m{W}$.

\item \label{fact_a3} A change of variables
$\phi^{\,'}_k=g_k(\theta_0,y^{k-1},a^k)+\phi_k$ reveals that the
inner sum does not depend on the value of
$g_k(\theta_0,y^{k-1},a^k)$.

\item \label{fact_a4} The dependence of the expression on $a_k$
has been removed.
\end{enumerate}

\addtocounter{enumi}{1}

\item By construction, $A_{k+1}^\infty$ is independent of
$(X^k,Y^k,A^k)$.\label{factD2}

\end{enumerate}

\end{proof}

Our horizon-free finite-alphabet universal scheme is now easily
adapted to use common randomness within the framework of this
section, as follows. First, active positions are removed (i.e.,
$b_a$=0). Instead, the type of each position and the repetition
position information for the update bits are directly provided by
the control sequence $A^\infty$. This is achieved (say) by using
an alphabet $\m{A}=\{training,regular\}\cup\NaturalF\cup\{0\}$,
where numerical values correspond to update positions and
determine which update bit is to be transmitted using which input
pair, taking the place of the $\Gamma^{M_u}$ described in
subsection \ref{subsec:modulo-additive}. Thus, $X_k$ is now
generated from $(\theta_0,Y^{k-1},A^k)$ instead of from
$(\theta_0,U^{k-1})$. Finally, the sequence of marginal
distributions $\m{P}$ is suitably defined taking into account the
removal of active positions (which can only improve the redundancy
term). Namely, for any position $j$ within the $k$th block we have
$P_j(training)=m_kb_k^{-1}$, $P_j(regular)=1-2m_kb_k^{-1}$, and a
uniform distribution over the numerical values $\brk{s_k}$ which
constitute the rest of the support of $P_j(\cdot)$ (where
$s_k=2|\m{X}|\lceil\log{(k+1)}\rceil$ corresponds to the number of
update bits, see (\ref{eq:upd_bit_num_horizon_free})). Given the
modifications described above, the adapted universal transmission
scheme under the new construction, either operating with or
without dithering, is well defined.

We are now ready to show that the adapted scheme with dithering
achieves the empirical capacity over $\m{C}_{\sstm{X}}$. Clearly,
the adapted scheme without dithering is essentially equivalent to
the scheme without common randomness discussed in previous
sections (up to the minor issue of active feedback replaced by
common randomness), and by repeating the same proof it is readily
verified that it achieves the empirical capacity over
$\mathscr{M}_{\sstm{X}}$ as well. Moreover, note that the fact
that $\m{W}\in\mathscr{M}_{\sstm{X}}$ was used in that proof
solely for the sake of Lemma \ref{lem:sampling}, namely to show
that the training pattern sequence and each of the update pattern
sequences, constitute causal sampling sequences for the noise
sequence within each block. Now, for a given channel
$\m{W}\in\mathscr{C}_{\sstm{X}}$ and a message point
$\theta_0\in[0,1)$, suppose the adapted scheme operates without
dithering over the corresponding induced channel
$\m{W}^\dagger(G,\m{P},\Delta,\m{W},\theta_0)$ with the same
message point $\theta_0$. In this case, Lemma
\ref{lem:sampling_dithering} verifies that under the construction
considered in this section, it still holds that the training and
update pattern sequences constitute causal sampling sequences for
the noise sequence within each block. Therefore, we conclude that
for any $\m{W}\in\mathscr{C}_{\sstm{X}}$ and $\theta_0\in[0,1)$,
the adapted scheme operating without dithering locally achieves
the empirical capacity for $(\m{W}^\dagger,\theta_0)$.
Furthermore, note that although the induced channel
$\m{W}^\dagger$ depends both on the message point and on the
channel $\m{W}$, the convergence parameters
$\eps_1(n),\eps_2(n),\eps_3(n)$ do not. Finally, according to
Lemma \ref{lem:equivalence}, the above implies that the adapted
scheme with dithering locally achieves the empirical capacity for
any pair of channel $\m{W}\in\mathscr{C}_{\sstm{X}}$ and message
point $\theta_0\in[0,1)$, with convergence parameters
$\eps_1(n),\eps_2(n),\eps_3(n)$ independent of $(\m{W},\theta_0)$.
Hence, by definition this scheme uniformly achieves the empirical
capacity over the family $\mathscr{C}_{\sstm{X}}$, and the proof
is concluded.

As discussed in section \ref{sec:result}, when operating over
$\mathscr{C}_{\sstm{X}}$ a uniform input distribution is essential
in order for the defined modulo-additive empirical capacity to be
meaningful, and in turn achievable. The discussion in this section
reveals the operational significance of this requirement within
the framework of our universal scheme. The entire scheme hinges on
the ability of a training sample to estimate the empirical
capacity of a block, and on the capability to reliably transmit
update information over any block whose empirical capacity is not
too small. Roughly speaking, a uniform input distribution
(obtained here by dithering) guarantees that with high
probability, the empirical distribution of the realized noise
sequence over an i.i.d. sample (i.e., training or update
positions) is close to that of the entire realized noise sequence.

\bibliographystyle{IEEEbib}
\bibliography{C:/Work/latex/ofer_refs_master}

\end{document}